\documentclass[conference,compsoc]{IEEEtran}
\usepackage{cite}

\usepackage[T1]{fontenc}
\usepackage[utf8]{inputenc}
\usepackage{pslatex}
\usepackage{xspace}
\usepackage{xurl}
\usepackage{dcolumn}
\usepackage[hidelinks]{hyperref}

\usepackage[kerning,spacing]{microtype} %
\usepackage{caption}
\usepackage{booktabs}
\usepackage{subcaption}
\usepackage{amsmath, txfonts, tikz, multirow}
\usepackage{pgfplots}
\usepackage{graphicx}
\usepackage{listings, listings-rust}
\usetikzlibrary{pgfplots.groupplots, matrix,fit}
\usetikzlibrary{calc,positioning,arrows}

\DeclareMathAlphabet{\mathcal}{OMS}{cmsy}{m}{n}
\DeclareMathAlphabet{\mathbf}{OT1}{cmr}{bx}{n}
\DeclareMathAlphabet{\mathsf}{OT1}{cmss}{m}{n}
\DeclareMathAlphabet{\mathit}{OT1}{cmr}{m}{it}
\DeclareMathAlphabet{\mathtt}{OT1}{cmtt}{m}{n}
\DeclareSymbolFont{symbols}     {OMS}{cmsy}{m}{n}
\DeclareSymbolFont{largesymbols}{OMX}{cmex}{m}{n}
\DeclareSymbolFont{letters}     {OML}{cmm} {m}{it}
\DeclareSymbolFont{operators}   {OT1}{cmr} {m}{n}

\usepackage{crypto-environments, setspace}

\usepackage{pifont}
\newcommand{\cmark}{\ding{51}}

\begin{document}

\date{}

\title{\Large \bf Private Hierarchical Governance for Encrypted Messaging}%

\author{
{\rm Armin Namavari$^{1}$, Barry Wang$^{2}$,
                Sanketh Menda$^{1}$, Ben Nassi$^{1}$,
                Nirvan Tyagi$^{3,4}$}\\
{\rm James Grimmelmann$^{2}$, Amy Zhang$^{3}$, Thomas Ristenpart$^{1}$}\\
{\rm $^{1}$ Cornell Tech \hspace*{3em}
$^{2}$ Cornell University \hspace*{3em}
$^{3}$ University of Washington} \hspace*{3em}
$^{4}$ Stanford University
}

\newcommand{\para}[1]{\noindent\textbf{#1}}

\renewcommand{\paragraph}[1]{\vspace*{4pt}\noindent\textbf{#1.}}

\def\authnote{1}

\newcounter{mynote}[section]
\definecolor{DeepPink}{HTML}{D60270}
\definecolor{DarkBlue}{HTML}{528AAE}
\newcommand{\notecolor}{blue}
\newcommand{\thenote}{\thesection.\arabic{mynote}}
\newcommand{\tnote}[1]{\ifnum\authnote=1\refstepcounter{mynote}{\bf \textcolor{\notecolor}{$\ll$TomR~\thenote: {\sf #1}$\gg$}}\fi}
\newcommand{\anote}[1]{\ifnum\authnote=1\refstepcounter{mynote}{\bf \textcolor{purple}{$\ll$ArminN~\thenote: {\sf #1}$\gg$}}\fi}
\newcommand{\ben}[1]{\ifnum\authnote=1\refstepcounter{mynote}{\bf \textcolor{purple}{$\ll$Ben~\thenote: {\sf #1}$\gg$}}\fi}
\newcommand{\sanketh}[1]{\ifnum\authnote=1\refstepcounter{mynote}{\bf \textcolor{DeepPink}{$\ll$sanketh~\thenote: {\sf #1}$\gg$}}\fi}
\newcommand{\amy}[1]{\ifnum\authnote=1\refstepcounter{mynote}{\bf \textcolor{brown}{$\ll$amy~\thenote: {\sf #1}$\gg$}}\fi}
\newcommand{\nirvan}[1]{\ifnum\authnote=1\refstepcounter{mynote}{\bf \textcolor{DarkBlue}{$\ll$Nirvan~\thenote: {\sf #1}$\gg$}}\fi}
\newcommand{\barry}[1]{\ifnum\authnote=1\refstepcounter{mynote}{\bf \textcolor{cyan}{$\ll$Barry~\thenote: {\sf #1}$\gg$}}\fi}

\newcommand{\fixme}[1]{\ifnum\authnote=1{\textcolor{red}{[FIXME: #1]}}\fi}
\newcommand{\fm}[1]{\ifnum\authnote=1{\textcolor{red}{\bf [#1]}}\fi}
\newcommand{\better}[1]{\ifnum\authnote=1{\textcolor{violet}{[BetterWord: #1]}}\fi}
\newcommand{\todo}[1]{\ifnum\authnote=1{\textcolor{red}{[TODO: #1]}}\fi}
\newcommand{\point}[1]{\ifnum\authnote=1{\textcolor{gray}{/* #1  */}}\fi}

\newcommand{\locClientCode}{1,431}
\newcommand{\locGovernanceLayer}{3,988}

\newcommand{\privGov}{Private Community Governance}
\newcommand{\privGovAcr}{PCG}

\newcommand{\cif}{\mathbf{if\;}}
\newcommand{\cthen}{\mathbf{\;then\;}}
\newcommand{\celse}{\mathbf{else\;}}
\newcommand{\creturn}{\mathbf{return\;}}
\newcommand{\ctrue}{\mathsf{true}}
\newcommand{\cfalse}{\mathsf{false}}
\newcommand{\cbad}{\mathsf{bad}}
\newcommand{\cflag}{\mathsf{flag}}
\newcommand{\sets}{\mathsf{sets\;}}

\newcommand{\indsm}{\hspace*{.75em}}
\newcommand{\indeqn}{\;\;\;\;\;\;\;}

\newcommand{\query}[1]{\procfont{query} {#1}:}
\newcommand{\queryl}[1]{\underline{\procfont{query} {#1}:}}
\newcommand{\procedure}[1]{\underline{\procfont{procedure} {#1}:}}
\newcommand{\procedurev}[1]{\underline{{#1}:}\smallskip}
\newcommand{\subroutine}[1]{\underline{\procfont{subroutine} {#1}:}}
\newcommand{\subroutinev}[1]{\underline{\procfont{subroutine} {#1}:}\smallskip}
\newcommand{\subroutinenl}[1]{{\procfont{subroutine} {#1}:}}
\newcommand{\subroutinenlv}[1]{{\procfont{subroutine} {#1}:}\smallskip}
\newcommand{\adversary}[1]{\underline{\procfont{adversary} {#1}:}}
\newcommand{\adversaryv}[1]{\underline{\procfont{adversary} {#1}:}\smallskip}
\newcommand{\experimentv}[1]{\underline{{#1}}\smallskip}
\newcommand{\algorithmv}[1]{\underline{\procfont{algorithm} {#1}:}\smallskip}

\newcommand{\sendOrderedMsg}{\mathsf{SendOrdMsg}}
\newcommand{\sendUnorderedMsg}{\mathsf{SendUnordMsg}}
\newcommand{\sendContentMsg}{\mathsf{SendContentMsg}}
\newcommand{\sendGovMsg}{\mathsf{SendGovMsg}}
\newcommand{\recvMsg}{\mathsf{RecvMsg}}
\newcommand{\sendReport}{\mathsf{SendReport}}
\newcommand{\verifReport}{\mathsf{VerifyReport}}
\newcommand{\inviteUser}{\mathsf{InviteUser}}
\newcommand{\addUser}{\mathsf{AddUser}}
\newcommand{\updateGroupState}{\mathsf{UpdateGroupState}}
\newcommand{\joinGroup}{\mathsf{JoinGroup}}
\newcommand{\accept}{\mathsf{Accept}}
\newcommand{\welcome}{\mathsf{welcomeMsg}}
\newcommand{\govSt}{\st_\textrm{gov}}
\newcommand{\mlsSt}{\st_\textrm{mls}}
\newcommand{\localSt}{\st_\textrm{loc}}
\newcommand{\conSt}{\st_\textrm{con}}
\newcommand{\mlsApi}{\mathsf{MLS}}
\newcommand{\policy}{P}
\newcommand{\cGov}{c_\textrm{gov}}
\newcommand{\exec}{\mathsf{Execute}}
\newcommand{\sk}{\mathit{sk}}
\newcommand{\pk}{\mathit{pk}}
\newcommand{\reportSk}{\sk_\mathrm{R}}
\newcommand{\reportPk}{\pk_\mathrm{R}}
\newcommand{\digSig}{\mathsf{S}}
\newcommand{\signMsg}{\mathsf{Sign}}
\newcommand{\verifyMsg}{\mathsf{Verify}}
\newcommand{\init}{\mathsf{init}}
\newcommand{\initClient}{\mathsf{InitClient}}
\newcommand{\createGroup}{\mathsf{CreateGroup}}
\newcommand{\gid}{\mathsf{gid}}
\newcommand{\report}{\rho}
\newcommand{\sendCommit}{\mathsf{SendCommit}}
\newcommand{\ordAppMsgProp}{\mathsf{OrdAppMsgProp}}

\newcommand{\send}{\mathsf{Send}}
\newcommand{\recv}{\mathsf{Recv}}
\newcommand{\sync}{\mathbf{Sync}}
\newcommand{\userIdLabel}{\mathsf{uid}}
\newcommand{\userIdVar}{\mathit{uid}}
\newcommand{\clientState}{C}
\newcommand{\roomId}{\mathit{rid}}
\newcommand{\messageVar}{\mathit{msg}}
\newcommand{\messageData}{\mathit{mdata}}
\newcommand{\messageId}{\mathit{mid}}
\newcommand{\textMsgStruct}{\mathsf{Text}} %
\newcommand{\textVar}{\mathit{txt}}
\newcommand{\textLabel}{\mathsf{txt}}
\newcommand{\refVar}{\mathit{ref}}
\newcommand{\pollStartMsgStruct}{\mathsf{PollStart}}
\newcommand{\optionVar}{\mathit{opt}}
\newcommand{\optionsVec}{\mathit{opts}}
\newcommand{\actionsVec}{\mathit{actns}}
\newcommand{\pollVoteMsgStruct}{\mathsf{PollVote}}
\newcommand{\pollEndMsgStruct}{\mathsf{PollEnd}}
\newcommand{\reactMsgStruct}{\mathsf{React}}
\newcommand{\reactionVar}{\mathit{rxn}}
\newcommand{\reportMsgStruct}{\mathsf{Report}}
\newcommand{\messagessVec}{\mathit{msgs}}
\newcommand{\reasonVar}{\mathit{rzn}}
\newcommand{\signature}{\sigma}
\newcommand{\escalate}{\mathsf{Escalate}}
\newcommand{\reportsVec}{\mathit{reports}}
\newcommand{\setStateMsgStruct}{\mathsf{SetState}}
\newcommand{\keyVar}{\mathit{key}}
\newcommand{\valVar}{\mathit{val}}
\newcommand{\changeNameStruct}{\mathsf{ChangeName}}
\newcommand{\nameVar}{\mathit{name}}
\newcommand{\changeTopicStruct}{\mathsf{ChangeTopic}}
\newcommand{\topicVar}{\mathit{topic}}
\newcommand{\blockUserStruct}{\mathsf{BlockUser}}
\newcommand{\setRoleMsgStruct}{\mathsf{DefRole}}
\newcommand{\roleVar}{\mathit{role}}
\newcommand{\typesVar}{\mathit{types}}
\newcommand{\setUserRoleMsgStruct}{\mathsf{SetUserRole}}
\newcommand{\kickUserMsgStruct}{\mathsf{KickUser}}
\newcommand{\inviteMsgStruct}{\mathsf{InviteUser}}
\newcommand{\removeMsgStruct}{\mathsf{RemoveMsg}}
\newcommand{\setTextFilterMsgStruct}{\mathsf{SetTextFilter}}
\newcommand{\filterVar}{\mathit{filter}}
\newcommand{\handleMessage}{\mathbf{HandleMsg}}
\newcommand{\handleKickUser}{\mathbf{HandleKickUser}}

\newcommand{\membersLabel}{\mathsf{members}}
\newcommand{\messagingProt}{\mathsf{MsgProt}}
\newcommand{\updateMembers}{\mathbf{UpdateMembers}}
\newcommand{\mainRoomLabel}{\mathsf{mainRoom}}

\newcommand{\messageSpace}{\mathsf{MsgSp}}
\newcommand{\messageIdSpace}{\mathsf{MsgIdSp}}
\newcommand{\textSpace}{\mathsf{TextSp}}
\newcommand{\pollSpace}{\mathsf{PollSp}}
\newcommand{\reactSpace}{\mathsf{ReactSp}}
\newcommand{\reportSpace}{\mathsf{ReportSp}}
\newcommand{\actionSpace}{\mathsf{ActionSp}}
\newcommand{\roomIdSpace}{\mathsf{RmIdSp}}
\newcommand{\userIdSpace}{\mathsf{UserIdSp}}
\newcommand{\clientStateSpace}{\mathsf{ClientStateSp}}
\newcommand{\messageTypes}{\mathsf{MsgTypes}}
\newcommand{\messageDataSpace}{\mathsf{MsgDataSp}}

\newcommand{\bits}{\{0,1\}}
\newcommand{\powerset}[1]{\mathcal{P}(#1)}
\newcommand{\getsr}{{\;{\leftarrow{\hspace*{-3pt}\raisebox{.75pt}{$\scriptscriptstyle\$$}}}\;}}
\newcommand{\getdist}[1]{{\;{\leftarrow{\hspace*{-3pt}\raisebox{.75pt}{$\scriptscriptstyle #1$}}}\;}}
\newcommand{\Z}{\mathbb{Z}}
\newcommand{\Zp}{\Z_p}
\newcommand{\pfRel}{\mathcal{R}}
\newcommand{\sigPok}{\mathsf{SPoK}}
\newcommand{\sigPokRel}{\sigPok^\pfRel}
\newcommand{\spokProve}{\mathsf{prove}}
\newcommand{\spokVerify}{\mathsf{verify}}
\newcommand{\zkPf}{\pi}

\newcommand{\myind}{\hspace*{5pt}}
\newcommand{\calP}{\mathcal{P}}
\newcommand{\hash}{H}
\newcommand{\sent}{\mathsf{sent}}
\newcommand{\recvFn}{R}
\newcommand{\advA}{\mathcal{A}}
\newcommand{\game}{\mathbf{G}}
\newcommand{\gameTranscriptConsistency}{\game^{\mathrm{tc}}}
\newcommand{\gameSenderBinding}{\game^{\mathrm{sb}}}
\newcommand{\aliceSend}{\mathbf{ASend}}
\newcommand{\bobSend}{\mathbf{BSend}}
\newcommand{\aliceRecv}{\mathbf{ARecv}}
\newcommand{\bobRecv}{\mathbf{BRecv}}
\newcommand{\aliceView}{\vec{v}_A}
\newcommand{\bobView}{\vec{v}_B}
\newcommand{\aliceViewCtr}{c_A}
\newcommand{\bobViewCtr}{c_B}
\newcommand{\aliceReportedView}{\vec{a}_v}
\newcommand{\bobReportedView}{\vec{b}_v}
\newcommand{\msg}{m}
\newcommand{\msgVec}{\vec{m}}
\newcommand{\msgMd}{M}
\newcommand{\msgMdVec}{\vec{M}}
\newcommand{\aliceQueue}{\vec{q}_A}
\newcommand{\bobQueue}{\vec{q}_B}
\newcommand{\aliceSentList}{A}
\newcommand{\bobSentList}{B}
\newcommand{\nonce}{\nu}
\newcommand{\nonceSpace}{\mathsf{NonceSp}}
\newcommand{\transcriptFrankingScheme}{\mathsf{TF}}
\newcommand{\frank}{\mathsf{Frank}}
\newcommand{\twoFrank}{\mathsf{TwoFrank}}
\newcommand{\forge}{\mathsf{Forge}}
\newcommand{\jforge}{\mathsf{JForge}}
\newcommand{\rforge}{\mathsf{RForge}}
\newcommand{\twoAMF}{\mathsf{2AMF}}
\newcommand{\groupAMF}{\mathsf{GAMF}}
\newcommand{\groupPk}{\pk_g}
\newcommand{\groupSk}{\sk_g}
\newcommand{\verify}{\mathsf{Verify}}
\newcommand{\judge}{\mathsf{Judge}}
\newcommand{\recon}{\mathsf{Recon}}
\newcommand{\frankMsg}{\mathsf{FrankMsg}}
\newcommand{\frankAck}{\mathsf{FrankAck}}
\newcommand{\verifyAck}{\mathsf{VerifyAck}}
\newcommand{\frankAckAck}{\mathsf{FrankAckAck}}
\newcommand{\verifyAckAck}{\mathsf{VerifyAckAck}}
\newcommand{\omb}{{1 - b}}
\newcommand{\amfSigVec}{\vec{\amfSig}}
\newcommand{\ackVec}{\vec{\ackMsg}}
\newcommand{\bitLabel}{\mathsf{bit}}
\newcommand{\judgeLevel}{\ell}
\newcommand{\self}{\mathsf{self}}
\newcommand{\hashLabel}{\mathsf{hash}}
\newcommand{\msgLabel}{\mathsf{msg}}
\newcommand{\recvHist}{\mathsf{recvHist}}
\newcommand{\sendHist}{\mathsf{sendHist}}
\newcommand{\msgs}{\mathsf{msgs}}
\newcommand{\acks}{\mathsf{acks}}
\newcommand{\amfSig}{\sigma}
\newcommand{\comMod}{c}
\newcommand{\platMod}{p}
\newcommand{\amfSigMod}{\amfSig_c}
\newcommand{\amfSigPlat}{\amfSig_p}
\newcommand{\keygen}{\mathsf{KeyGen}}
\newcommand{\senderPk}{\pk_s}
\newcommand{\senderSk}{\sk_s}
\newcommand{\recPk}{\pk_r}
\newcommand{\recSk}{\sk_r}
\newcommand{\judgePk}{\pk_j}
\newcommand{\judgeSk}{\sk_j}
\newcommand{\judgePkZ}{\pk_0}
\newcommand{\judgePkO}{\pk_1}
\newcommand{\judgeSkZ}{\sk_0}
\newcommand{\judgeSkO}{\sk_1}
\newcommand{\judgeSt}{\st_j}
\newcommand{\alicePk}{\pk_A}
\newcommand{\aliceSk}{\sk_A}
\newcommand{\aliceSt}{\st_A}
\newcommand{\bobPk}{\pk_B}
\newcommand{\bobSk}{\sk_B}
\newcommand{\bobSt}{\st_B}
\newcommand{\alicePerm}{\pi_A}
\newcommand{\bobPerm}{\pi_B}
\newcommand{\senderPerm}{\pi_s}
\newcommand{\recvPerm}{\pi_r}
\newcommand{\st}{\mathit{st}}
\newcommand{\senderSt}{\st_s}
\newcommand{\recSt}{\st_r}
\newcommand{\reportInfoVec}{\vec{\reportInfo}}
\newcommand{\reportInfo}{\rho}
\newcommand{\reportOutput}{R}
\newcommand{\pkLabel}{\mathsf{pk}}
\newcommand{\skLabel}{\mathsf{sk}}
\newcommand{\validFlag}{\mathit{valid}}
\newcommand{\platformView}{\vec{v}_P}
\newcommand{\platformCtr}{i}
\newcommand{\ctrLabel}{\mathsf{ctr}}
\newcommand{\TCAdv}{\mathbf{Adv}^{\mathrm{tc}}_\transcriptFrankingScheme}
\newcommand{\SBAdv}{\mathbf{Adv}^{\mathrm{sb}}_\transcriptFrankingScheme}
\newcommand{\amfScheme}{\mathsf{AMF}}
\newcommand{\tfSig}{\sigma_\transcriptFrankingScheme}
\newcommand{\lastMsgHash}{\mathsf{lastMsgH}}
\newcommand{\lastAckHash}{\mathsf{lastAckH}}
\newcommand{\lastSentHashLabel}{\mathsf{lastSentHash}}
\newcommand{\lastSentHash}{h_s}
\newcommand{\lastRecvHash}{h_r}
\newcommand{\lastRecvHashLabel}{\mathsf{lastRecvHash}}
\newcommand{\sentMsgLabel}{\mathsf{sent}}
\newcommand{\recvMsgLabel}{\mathsf{recv}}
\newcommand{\append}{\mathsf{Append}}
\newcommand{\senderMsgs}{\msgMdVec_s}
\newcommand{\recvMsgs}{\msgMdVec_r}
\newcommand{\senderSigs}{\vec{\sigma}_s}
\newcommand{\recSigs}{\vec{\sigma}_r}
\newcommand{\lastHash}{\hat{h}}
\newcommand{\lastMsgLabel}{\mathsf{lastMsg}}
\newcommand{\lastHashLabel}{\mathsf{lastHash}}
\newcommand{\ackCtr}{i}
\newcommand{\recAckCtr}{i_r}
\newcommand{\senderAckCtr}{i_s}
\newcommand{\aliceAckCtr}{i_A}
\newcommand{\bobAckCtr}{i_B}
\newcommand{\aliceSendCtr}{s_A}
\newcommand{\bobSendCtr}{s_B}
\newcommand{\ackMsg}{\alpha}
\newcommand{\msgNum}{\mathsf{msgCtr}}
\newcommand{\lastAckNum}{\mathsf{lastAckNum}}
\newcommand{\ackNum}{\mathsf{ackCtr}}
\newcommand{\ackMsgNum}{\mathsf{ackMsgNum}}
\newcommand{\ackStruct}{\mathsf{Ack}}
\newcommand{\msgStruct}{\mathsf{Msg}}
\newcommand{\ackOrMsg}{\mu}
\newcommand{\ackOrMsgVec}{\vec{\ackOrMsg}}
\newcommand{\lastSentNum}{\mathsf{lastSentNum}}
\newcommand{\alice}{A}
\newcommand{\bob}{B}

\newcommand{\reconViewBob}{\mathbf{BReconView}}
\newcommand{\reconViewAlice}{\mathbf{AReconView}}
\newcommand{\reconViewSender}{\mathbf{SReconView}}
\newcommand{\reconViewReceiver}{\mathbf{RReconView}}
\newcommand{\reconView}{\mathbf{ReconView}}
\newcommand{\vecView}{\vec{v}}
\newcommand{\currVert}{c}
\newcommand{\inwardVertFn}{\mathbf{In}}
\newcommand{\aliceReportVec}{\vec{a}_r}
\newcommand{\aliceElem}{a}
\newcommand{\bobElem}{b}
\newcommand{\bobReportVec}{\vec{b}_r}
\newcommand{\emptyVec}{()}
\newcommand{\messageDAG}{G}
\newcommand{\verifyChain}{\mathbf{VerifyChain}}
\newcommand{\verifyReportBob}{\mathbf{VerifyReportBob}}
\newcommand{\concat}{\mathbf{Concat}}

\newcommand{\user}{U}
\newcommand{\usermod}{\hat{U}}
\newcommand{\platmod}{\hat{P}}
\newcommand{\govAction}{A}
\newcommand{\govActionType}{\alpha}
\newcommand{\kvKey}{K}
\newcommand{\kvValue}{V}
\newcommand{\permissionLevel}{P}

\newcommand{\Colon}{\::\:}
\newcommand{\sends}[3]{#1 \rightarrow #2\Colon #3}
\newcommand{\platreports}[3]{#1 \rightsquigarrow #2\Colon #3}
\newcommand{\reports}[3]{#1 \Rightarrow #2\Colon #3}
\newcommand{\initvote}[2]{#1 \textnormal{ noms } #2}
\newcommand{\votes}[2]{#1 \textsf{ votes } #2}

\newcommand{\vecM}{\vec{M}}

\newcommand{\msgtriple}[3]{(#1\rightarrow#2:#3)}
\newcommand{\adds}[2]{#1 \textsf{ adds } #2}
\newcommand{\kicks}[2]{#1 \textsf{ kicks } #2}
\newcommand{\timeout}[2]{#1 \textsf{ timeouts } #2}
\newcommand{\removes}[2]{#1 \textsf{ rems } #2}
\newcommand{\blocks}[2]{#1 \textsf{ blocks } #2}
\newcommand{\unblocks}[2]{#1 \textsf{ okays } #2}
\newcommand{\appoints}[2]{#1 \textsf{ appoints } #2}
\newcommand{\setsKV}[3]{#1 \textsf{ sets } (#2, #3)}
\newcommand{\givesPerm}[3]{#1 \textsf{ permits } (#2, #3)}
\newcommand{\setsActionPerm}[3]{#1 \textsf{ requires } (#2, #3)}

\newcommand{\sendUam}{\textsf{send\_uam}}
\newcommand{\sendOam}{\textsf{send\_oam}}
\newcommand{\add}{\textsf{add\_user}}
\newcommand{\remove}{\textsf{remove\_user}}
\newcommand{\update}{\textsf{update\_user}}
\newcommand{\getepoch}{\textsf{get\_epoch}}

\newcommand{\sysname}{MlsGov\xspace}
\newcommand{\baseline}{MlsBase\xspace}

\newcommand{\filterproc}{\texttt{filter}\xspace}
\newcommand{\initproc}{\texttt{init}\xspace}
\newcommand{\checkproc}{\texttt{check}\xspace}
\newcommand{\passproc}{\texttt{pass}\xspace}
\newcommand{\failproc}{\texttt{fail}\xspace}

\setlength{\tabcolsep}{2pt}

\maketitle
\begin{abstract}

    The increasing harms caused by hate, harassment, and other forms of abuse
    online have motivated major platforms to explore hierarchical governance.
    The idea is to allow communities to have designated members take on
    moderation and leadership duties; meanwhile, members can still escalate
    issues to the platform. But these promising approaches have only been
    explored in plaintext settings where community content is public to the
    platform. It is unclear how one can realize hierarchical governance in the
    huge and increasing number of online communities that utilize end-to-end
    encrypted (E2EE) messaging for privacy.

    We propose private hierarchical governance systems. These
    should enable similar levels of community governance as in plaintext
    settings, while maintaining cryptographic privacy of content and
    governance actions not reported
    to the platform. We design the first such system, taking a layered approach
    that adds governance logic on top of an encrypted messaging protocol; we
    show how an extension to the message layer security (MLS) protocol suffices
    for achieving a rich set of governance policies. Our approach allows
    developers to rapidly prototype new governance features, taking inspiration
    from a plaintext system called PolicyKit. We build a prototype
    E2EE messaging system called \sysname that supports content-based
    community and platform moderation, elections of community moderators, votes
    to remove abusive users, and more.
\end{abstract}

\section{Introduction}
\label{sec:intro}

Today, end-to-end encryption (E2EE) helps protect the private
communications of billions of people~\cite{whatsapp-2b} from data breaches,
overzealous advertisers, and snooping governments. At the same time,
surveys~\cite{sokAbuse,pew21} show that people are being harmed by online abuse:
almost half of respondents report experiencing abuse online, the fraction of
those reporting experiencing abuse is growing each year, and much of this abuse
is carried out over messaging apps.

As a result, many large E2EE messaging platforms have dedicated trust and safety
teams that develop and execute moderation policies to mitigate abuse. But E2EE
makes moderation hard with existing techniques~\cite{SchefflerM23}.  Content-oblivious
approaches~\cite{ContentOblivious} do not handle most types of abuse, and those
that allow reporting
content~\cite{whatsapp-reporting,MessengerSecretConvo,SMFCommitting,SMFFast,AMF,Hecate}
rely on a centralized platform-operated moderation service that, in turn, relies
on a combination of automated tools and large groups of human
moderators~\cite{crawford2016flag, AutoContentModerationScale, klonick2017new,
  grimmelmann2015virtues}. Such centralized moderation infrastructure represents a
dangerous accumulation of power~\cite{NetworkGatekeeping, suzor2019lawless} and
struggles with nuanced, contextualized, or community-specific
abuse~\cite{propublica-whatsapp}.

Some plaintext platforms, like Reddit \cite{RedditModeration} and
Discord~\cite{DiscordModeration, jiang2019moderation} instead employ a
hierarchical governance structure. Certain community members, known as
\textit{community moderators}, define and enforce community policies, while
moderators at the platform level oversee communities and enforce platform
policies. This leads to a separation of powers in which most moderation of
user activity happens at the community level. Community moderators also
benefit from tools that automate parts of the moderation process
\cite{jhaver2019human}. For instance, Subreddit moderators can
automatically flag posts based on keywords with the AutoModerator tool,
lessening the burden of moderation.
Decentralizing and distributing power in this way can better respect user agency
and enables a diversity of community approaches to governance
\cite{MultiLevelGov}, and much work has explored the design of tools to support
or enable community
governance~\cite{policykit,jhaver2019human,kiene2020uses,geiger2016bot} for
plaintext systems.

In this paper, we explore for the first time \emph{private hierarchical
  governance} for encrypted group messaging. This entails a platform design
that provides rich community-level governance features, while achieving
strong E2EE privacy, integrity, and accountability guarantees for both
content and governance-related tasks. For example, even a malicious
platform should not be able to infer who are a community's moderators,
nor undetectably interfere with governance actions moderators have taken.
As in the plaintext setting, we aim to provide community moderators with
tools that help automate moderation tasks, like Reddit's AutoModerator.
We give users the choice of reporting content to community moderators and
platform moderators. Given the private nature of the E2EE setting, the
platform will mainly rely on user reports to inform its moderation.

Realizing private hierarchical governance in the E2EE setting requires
overcoming a number of challenges, chief among them that the platform is
assumed to be malicious and must not learn anything about content exchanged
within groups. This complicates enabling automated moderation policies,
similar to those available to community moderators on Reddit and Discord,
in which the platform handles policy execution. One could enable such
policies by adding a centralized governance service as a trusted endpoint,
however this is undesirable from a usability and security standpoint. Users
should not be burdened with setting up the infrastructure for such an
endpoint nor should they have to trust a third-party platform for managing
one.

Our approach instead shifts community governance to be client-side logic.
But in so doing, we must determine how to manage governance actions in a
distributed, asynchronous network setting where the messaging platform is
potentially untrustworthy. One straw approach would be to just use standard
techniques for consensus~\cite{pease1980reaching,ongaro2014search} or
state-machine replication~\cite{schneider1990implementing} to all community
content and actions, but this would not be practical. A key enabling
insight is that our desired governance functions can be achieved while only
requiring clients to consistently agree on a small portion of community
state related to governance tasks, allowing more expensive consensus
mechanisms to only be needed rarely.

Given this insight, we detail a modular approach to private, hierarchical
governance. We suggest extending E2EE group messaging systems to support
ordered application messages (OAMs) for which the group achieves consensus
on the content and order of these messages. Existing E2EE protocols such as
message layer security (MLS)~\cite{mls-protocol,mls-architecture} already
have suitable consensus mechanisms to support our OAM extension; our
suggested extension to MLS can be viewed as a generalization of a recent
mechanism proposed in~\cite{cryptoAdmin}. We expect our extension to be of
broad utility; here we show how we can build a governance layer distributed
across clients in a group that agree on state via OAMs.

Our governance layer provides a full role-based access control
(RBAC)~\cite{rbac} mechanism and a policy engine, inspired by prior work
\cite{policykit}, that allows execution of expressive policies securely
using OAMs. To demonstrate the generality of our governance framework, we
use it to implement a policy for voting on changing the group name. We
analyze how our approach provides strong cryptographic guarantees of
governance privacy, integrity, and accountability that prevents adversarial
platforms from monitoring or interfering with community messages or
governance actions, and prevents adversarial clients from preventing abuse
reports or reporting messages they did not receive. We also support
reporting abuse to platform moderators to enable hierarchical governance;
unreported messages stay private from them.
Note that our focus is on providing the technical infrastructure for
realizing a broad range of automated governance policies, and leave to future work how to guide the design of
good policies and prevent abusive policies.

In order to provide a concrete instantiation of private hierarchical
governance, we build a full prototype E2EE messaging platform called
\sysname on top of a suitably modified MLS implementation, and show via
extensive performance analysis that our governance framework is practical.
To encourage further work on building governance for encrypted messaging,
we release \sysname as an open-source project\footnote{\url{https://github.com/AME2E/MLSGov}}.

We summarize our contributions below:

\begin{newitemize}
  \item  Our paper proposes the goal of private, hierarchical governance,
  which shares moderation and other tasks across communities and platforms,
  while retaining the security benefits of E2EE.
  \item We present a design that allows for governance policies while
  limiting the amount of consensus needed across clients. An extension to
  MLS that we propose allows for shared governance state. We analyze the
  security of our design by reasoning through possible attacks.
  \item  To demonstrate practicality, we build a prototype E2EE messaging
  platform called \sysname that is the first to support a wide range of
  governance features previously only available in plaintext settings, and
  measure its performance.
\end{newitemize}

\section{Background and Related Work}
\label{sec:relwork}

\begin{figure}[t]
    \center
    \scriptsize

    \begin{tabular}{lllllll}
        \textbf{System}                                                      & \textbf{\begin{tabular}[c]{@{}l@{}}E2EE \\
                                                                                               Content\end{tabular}} & \textbf{\begin{tabular}[c]{@{}l@{}}Gov. \\
                                                                                                                               Confidentiality\end{tabular}} & \textbf{\begin{tabular}[c]{@{}l@{}}Gov. \\
                                                                                                                                                                       Integrity\end{tabular}} & \textbf{RBAC} &
        \textbf{\begin{tabular}[c]{@{}l@{}}Generic \\ Policies\end{tabular}} &
        \textbf{\begin{tabular}[c]{@{}l@{}}Gov.    \\ Enforcement\end{tabular}}                                                                                                                                              \\
        \midrule
        Matrix                                                               & Yes
                                                                             & No                                         & No
                                                                             & Multi-role                                 & No
                                                                             & Server
        \\ \midrule
        Signal                                                               & Yes
                                                                             & Yes                                        & No
                                                                             & Two-role                                   & No
                                                                             & Server
        \\ \midrule
        Whatsapp                                                             & Yes
                                                                             & No*                                        & No*
                                                                             & Two-role                                   & No
                                                                             & Server*
        \\ \midrule
        Discord                                                              & No                                         & No
                                                                             & No                                         & Multi-role                                 & Yes                                        & Server \\ \midrule
        A-GCKA                                                               & Yes
                                                                             & No                                         & Yes
                                                                             & Two-role                                   & No
                                                                             & Client
        \\ \midrule
        \textbf{This Work}                                                   & \textbf{Yes}                               & \textbf{Yes}
                                                                             & \textbf{Yes}                               &
        \textbf{Multi-role}                                                  & \textbf{Yes}                               & \textbf{Client}                                                                                  \\ \midrule
    \end{tabular}

    \caption{A comparison between our work and other messaging platforms
        that support communities. * - based on our assessment (no public documentation).}
    \label{fig:relwork-table}
\end{figure}

\paragraph{Community governance}
Social media platforms have developed a wide range of governance strategies
to counter abusive content,  including both industrial moderation
approaches as well as community-reliant
approaches~\cite{caplan2018content,gillespie2018custodians}. In industrial
approaches, governance is primarily centralized at the platform level, with
policies set by trust and safety teams and carried out by commercial
content moderators~\cite{klonick2017new,roberts2019behind}, whereas in
community-reliant platforms, communities have broader latitude to define
their own governance. In these cases, each community has volunteer
moderators who can designate and enforce community-specific
rules~\cite{matias2019civic,seering2019moderator}. However, the platform
still retains the power to step in and enforce its rules, including banning
accounts and removing or quarantining entire
communities~\cite{RedditQuarantine,chandrasekharan2017you}, forming a
two-level hierarchical governance structure \cite{MultiLevelGov}.

Some community-reliant platforms provide a powerful set of tools to
communities to  automatically carry out
policies~\cite{seering2019moderator}. For instance, subreddit moderators
can use the Automoderator tool to automatically filter and act on posts
based on their content \cite{jhaver2019human}. Discord has a strong
ecosystem of third-party tools that community moderators can use to
customize their community \cite{kiene2020uses}. More recently, academic
work has explored richer governance approaches, where communities can
define and execute arbitrarily complex governance procedures written in
code \cite{policykit}, enabling communities to vote on new
moderators, enforce content filters, or institute reputation systems.

\paragraph{Abuse mitigations in E2EE settings}
In E2EE messaging platforms, governance is complicated by the fact that
moderators cannot by default verify the plaintext contents of
communications. A body of work has explored how to enable cryptographically
secure reporting of individual encrypted messages, starting
with Facebook's  message franking feature~\cite{MessengerSecretConvo}.
Subsequent academic work formalized message franking and characterized
(in)secure methods for achieving it~\cite{SMFCommitting,SMFFast, AMF,
    Hecate}. In addition to reporting abusive messages, moderators may benefit
from the ability to trace the spread and identify the origin of forwarded
messages. Recent work has explored how to realize this feature for E2EE
platforms~\cite{Traceback, SourceTracking, Hecate}.

Another abuse mitigation approach is automated content detection. Given the
scale at which content is exchanged on platforms, many plaintext platforms
have turned to automated content detection that alerts platforms when
particular content is sent or received~\cite{levy2022thoughts}. Perceptual
hashing~\cite{PerceptualHashing} is a popular tool used in these
approaches, particularly for the detection of child sexual abuse media
(CSAM). These techniques are not immediately applicable to E2EE platforms
as platform servers cannot observe the plaintext contents of messages and
client devices might not be allowed to see the plaintext contents of the
harmful content list (especially in the case of CSAM).  Recent work has
explored how multi-party computation can be used to navigate these
challenges while respecting user privacy~\cite{IdHarmfulMedia, ApplePSI}.
Such proposals have been met with criticism owing to their potential to
undermine the privacy goals of E2EE~\cite{BugsPockets} because they
automate reporting to platform moderators without user consent.

\paragraph{Existing E2EE governance tools}
Given the difficulty of conducting effective moderation at the platform
level without violating privacy, some attention has been paid toward
strengthening governance tooling at the community level. Matrix
\cite{MatrixSpec}, a protocol for federated E2EE messaging, provides some
basic community moderation features via role-based access control (RBAC)
\cite{MatrixModeration}. However, information about user roles is directly
visible to the homeserver (platform server). Mjolnir \cite{MatrixMjolnir}
provides server-level moderation features such as banning users, taking
down content, and managing user accounts. However, it does not enable
arbitrary programmable governance and cannot process content in encrypted
rooms.

WhatsApp's recent launch of Communities makes it easier for groups of over
a thousand people to use the platform~\cite{WhatsappCommunities}. Whatsapp
groups allow for two-level access control (between users and moderators).
Although no public documentation describes how or where Whatsapp enforces
governance, it bears mentioning that group metadata such as the profile
picture and topic associated with a group are visible to Whatsapp. As such,
we suspect that governance metadata such as the roles and permissions
within a group are visible to the platform and that the platform
enforces these roles.

Signal has explored private group management for E2EE chats, using a
combination of anonymous credentials and a server-managed encrypted list of
members~\cite{SignalPermissions}. Their design allows the server to
authenticate community moderators in the sealed sender setting. This
achieves governance confidentiality, but the reliance on platform-enforced
access control weakens governance integrity. There is also some leakage of
which encrypted entries perform actions on the governance state.
Furthermore, their design does not handle other aspects of governance, such
as allowing for rich and programmable moderation policies, such as those
enabled via Discord moderation bots, the Reddit Automoderator tool, and
PolicyKit \cite{policykit}.

The recently proposed A-CGKA construction~\cite{cryptoAdmin} suggests
embellishments to the message layer security (MLS) protocol involving
cryptographically enforced roles, such as administrators, for group
encrypted messaging. Their solution, however, does not describe a mechanism
for keeping administrator roles confidential. They achieve governance
integrity, but do not tackle governance confidentiality as a goal. While
they do not provide the rich governance features our solution provides, our
modifications to MLS are similar to theirs; we compare and contrast in more
detail in \secref{sec:extmls}.

These solutions all stop short of the rich governance features that many
communities have available to them on non-E2EE community-reliant platforms.
In addition, there has been little work examining how community-level
governance intersects with governance happening at the platform level, such
as banning accounts and communities from the platform.
We summarize the governance capabilities of popular messaging platforms in
\figref{fig:relwork-table}.  We define governance privacy and integrity in
detail \secref{sec:setting}. Role based access control (RBAC) allows
users to define permissions and roles within their community. The
``Generic Policies'' column refers to whether the system supports
programmable governance logic. We also indicate whether governance is
enforced on the platform server or on client devices. Asterisks
indicate aspects that are not publicly documented and the values we
provide in those entries are our conjectures based on the available
information we have.
In summary, no prior work tackles the design of rich governance tools in
E2EE that takes into account both community and platform levels. Our work
fills this gap.

\section{Overview and Goals}
\label{sec:setting}

Our paper presents a framework for building and testing governance for E2EE
online communities. In this section we provide an overview of our goals and
approach, and detail various components further in subsequent sections.

\paragraph{Private, hierarchical governance}
We seek to enable governance frameworks for E2EE social media, in
particular private direct and group messaging like that offered by Signal
and WhatsApp. But unlike today's messaging solutions, our system should
allow groups of users to form self-governed communities with the support of
a rich suite of governance features. Going forward, we use the term
``group'' and ``community'' interchangeably. A group should be able to
elect one or more moderators who retain special privileges within the group
to perform various actions, including removing users, adding users, and
blocking messages. These moderators can change over time. Group members
should be able to send abuse reports to community moderators to help inform
moderation decisions. We further aim to provide group members with features
that help automate parts of governance, such as handling votes for new
moderators or enforcing a moderator-specified word filter.

\begin{figure}[t]
    \center
    \includegraphics[width=0.7\linewidth]{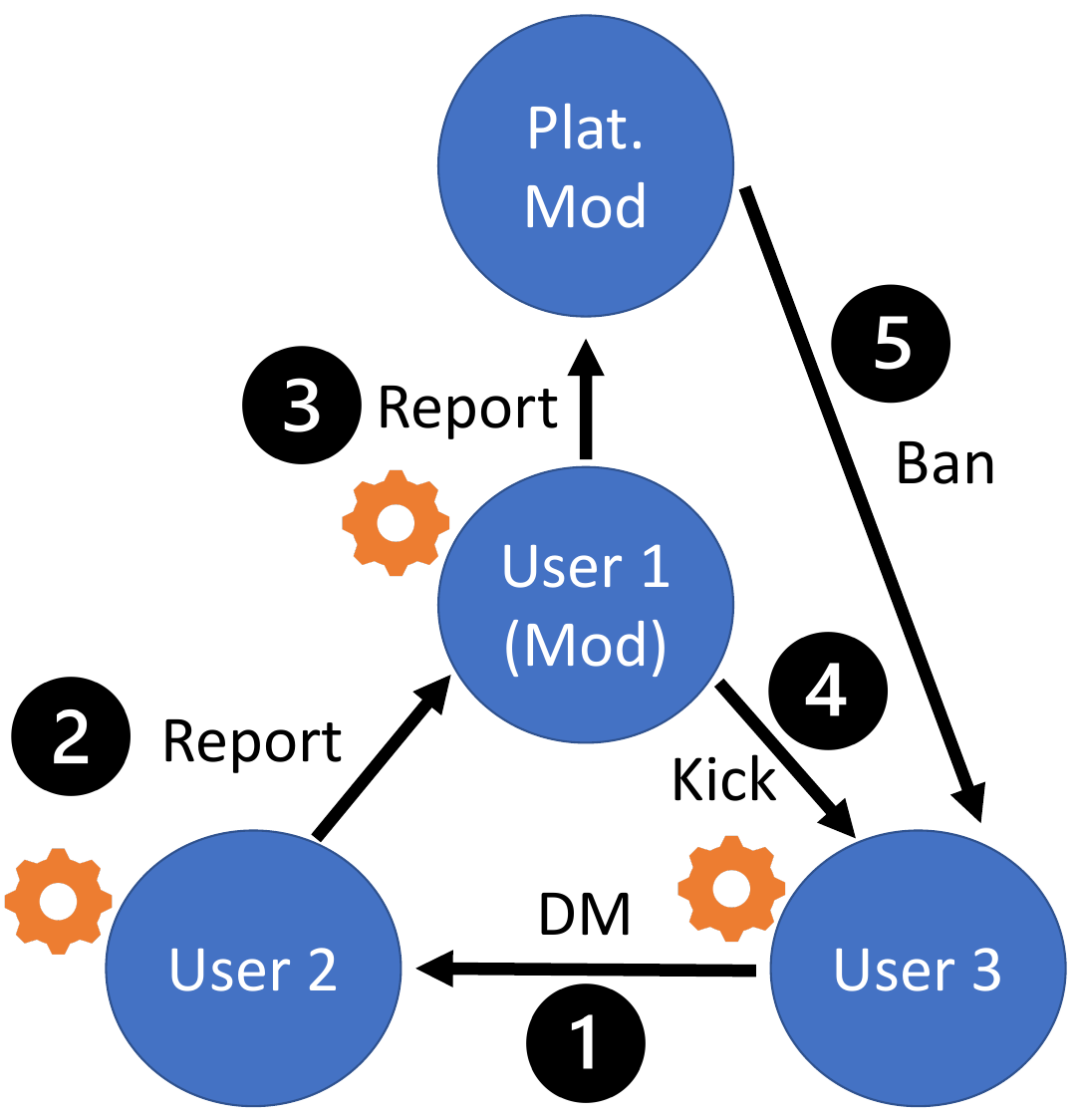}

    \caption{Diagram of an example governance scenario in which a user
        $\user_2$ reports another user $\user_3$ to a community
        moderator~$\user_1$, who kicks that user from the community and escalates
        by reporting the abuse to the platform's moderation systems. This results
        in a temporary ban of $\user_3$ from the platform. Execution of community
        governance is handled on clients (denoted by the gears) and remains
        private to the community. Only when a user chooses to report to the
        platform does the latter become involved.}
    \label{fig:diagram}
\end{figure}

Consider the following scenario, which we will use as a running example to
illustrate the types of governance issues that our framework can handle. A
community consisting of $N$ users $U_1, \ldots, U_N$ has a single
moderator~$\user_1$. Group member~$\user_2$ proposes to change the
community guidelines so that profanity is not allowed within the group. The
other users vote to pass this change.
Later,~$\user_3$ sends an abusive direct message (DM) to~$\user_2$ for
proposing this change. After receiving the message,~$\user_2$ reports the
message to the moderator~$\user_1$, who decides to remove~$\user_3$ from
the community as a result. As~$\user_1$ knows sending abusive messages is
against platform guidelines,~$\user_1$ forwards the report to the platform
moderation endpoint~$M$. A platform moderator receives the report and
decides to ban~$\user_3$ for one week for violating platform guidelines. As
a result~$\user_3$ is temporarily unable to participate in any community
hosted on the platform.
A diagram summarizing this scenario appears in
\figref{fig:diagram}.

Looking ahead to our threat model, we want governance actions like those in
the scenario above to be \emph{private} by default. That means that
governance actions should be cryptographically secure and opaque to the
messaging platform provider (from now on, simply the platform), so that,
for example, the platform should not need to know that~$\user_1$ is the
moderator, that~$\user_2$ proposed a new policy, that it was adopted and
who voted for it, or that $\user_3$ violated it.  Designing privacy as a
default for governance is conservative: not all users and communities
require such privacy, but it can be critical for those that do---e.g.,
political activists, journalists, or community organizers from marginalized
groups, any of whom may be targeted by nation-state or other powerful
adversaries. In particular, governance metadata can be sensitive
information; for instance, being a moderator of an activist group may also
indicate being a leader within that group.

But not all governance actions can be handled within a single community.
For example, $\user_1$ decided in the scenario that the in-community
problems caused by~$\user_3$ warrants further action, possibly to help
protect other communities from~$\user_3$ misbehavior. Thus, we want to
balance privacy with accountability and support what we call
\emph{hierarchical governance}. In addition to group-level governance,
users can choose to report in-group misbehavior, as done by~$\user_1$ in
the example. Even abusive behaviors of community moderators should be
reportable. Because the platform ultimately retains control over who is
allowed to use it, governance forms a hierarchy where platform moderators
can overrule community moderators.

Such hierarchical governance has already been explored in plaintext systems
such as subreddits and Facebook Groups~\cite{MultiLevelGov},
but adding in privacy surfaces new challenges, since by default platform
moderators will not have access to group messages or community moderator
identities. A key challenge addressed by our work is providing a framework
to explore the design space of private, hierarchical governance policies
and supporting mechanisms.

\paragraph{Platform architecture and layered design}
To achieve E2EE communities with private, hierarchical governance, we will
use an architecture consisting of three services: a \emph{delivery service}
(DS), an \emph{authentication service} (AS), and a \emph{platform
    moderation service}~(MS).

The DS and AS form what we call the \emph{messaging layer}.  The DS
provides API endpoints to route messages and service metadata between
clients. The AS provides bindings between identities and public keys.
These are similar to existing messaging service solutions and serve to
ensure confidential and authenticated delivery of messages between users
intermediated by a platform. In particular, the DS and AS are analogous to
services of the same name in the ongoing MLS~\cite{mls-architecture}
standard. However, governance needs cryptographic support from the
messaging layer, in particular the ability of a group of clients to
asynchronously agree on an ordered sequence of governance messages (in the
presence of potentially adversarial clients). We show how a relatively
straightforward modification to MLS provides the features necessary to
build our governance layer in a modular way (see \secref{sec:extmls}).

Logically sitting above the messaging layer we have a \emph{governance
    layer}. This layer consists of logic within the clients to enact governance
actions taken within the community, such as group members electing a new
moderator or a group moderator kicking a member. One approach here would be
to hard-code particular governance mechanisms, but we instead adopt the
viewpoint underlying PolicyKit~\cite{policykit} that governance should be
easily customizable. PolicyKit allows on-the-fly customization of the
software used by individual communities on a platform. We build a similarly
expressive governance layer for the E2EE setting. To address our security
goals (detailed below), the policy engine execution is handled by clients
that must agree on the current state of the group. Managing this state in
the presence of adversarial clients introduces complexity compared to
insecure plaintext approaches, but our layered design approach means that
the complexity can be largely ignored by those developing applications on
top of our governance layer. We describe the governance layer design in
\secref{sec:design}.

To provide hierarchical governance, we include the MS to receive abuse
reports from community members. For simplicity, we realize the MS via a
distinguished, virtual user (operated by the platform or someone to which
it delegates) to which report messages can be sent---this ensures a level
of simplicity and flexibility that will be beneficial in deployment. We
emphasize that the MS only becomes involved when a user affirmatively
chooses to submit a report to the platform moderator: in our example above
the MS only observes the community moderator's report about~$\user_3$.

Lastly, all of this enables the \emph{application layer} in which
applications that support community governance may be built. As a showcase,
we build a group messaging application using \sysname that supports
expressive governance policies.

\paragraph{Threat model and security goals}
Our system is designed to achieve our governance and privacy goals even in
the presence of malicious parties. We use the term malicious to refer to
parties that deviate from the protocol, and use the term abusive to refer
to clients that send legitimate protocol messages but with an intent to
cause harm.  We consider both malicious users within the community and a
malicious platform-operated DS and MS. Our design assumes that the AS is
honest, which can be enforced using separate mechanisms such as public key
infrastructure (PKI) transparency~\cite{CONIKS,SEEMless,VeRSA,Mog}.
In accordance with the MLS Architecture Document \cite{mls-architecture} \S
2.3.3, we expect the DS to aid with message ordering when needed and to
allow for eventual delivery of client messages sent over successful network
connections. Nonetheless, when the DS violates these expectations, we will
have mechanisms for the eventual detection of such misbehavior.

\textbf{\emph{Integrity:}}
Recall that we introduce a governance layer that includes client-side
state, called the governance state.  State will include information like
the permissions of group members such as who is a moderator, what policies
are being enforced, etc. As such governance state evolves over time and the
current governance state determines how actions performed by users produces
future versions of the governance state. A system achieves \emph{governance
    integrity} if all honest, online clients in a group agree on the same
sequence of governance states. This requires that clients observe a
consistent sequence of governance state updates and that they apply these
updates according to their governance functionality.

We target achieving governance integrity even in the presence of one or
more malicious clients in the group that collude with a malicious DS and
MS.
A malicious DS can drop and reorder encrypted messages, including
governance-related messages. This means that denial of service by a
malicious DS is always possible. A malicious DS can partition the group
into subgroups, by not delivering messages across their maliciously chosen
partition. This can fork the governance state across the two (or more)
subgroups, but such an attack should be detected as soon as the first
cross-partition message is delivered. We note that this is the same
security level achieved by MLS (c.f.,~\cite{mls-architecture}), and absent
further infrastructure there is no way to avoid partitioning.

\textbf{\emph{Accountability:}} We target \emph{user and reporter
    accountability}. This means that abusive users, even ones that are
malicious (i.e., using compromised, adversarial client software) should be
reportable to honest community moderators and to the MS. Reporting an
abusive user to an abusive community moderator may not lead to desired
outcomes. So user accountability extends to moderators, meaning any user
should be able to report abusive moderators to the MS. On the flipside,
reporter accountability means that community moderators and the MS must
reject malicious reports, such as reporting messages that were never sent.
In the cryptographic literature~\cite{AMF}, user and reporter
accountability are sometimes referred to as sender and receiver binding,
respectively.

An abusive group is one in which all members are abusive. For example, a
group that is trading child-sexual abuse material (CSAM) or other abusive
content. In this work, we have as a non-goal automated detection of abusive
communities, such as considered in recent proposals by
Apple~\cite{ApplePSI} and others~\cite{IdHarmfulMedia}. While automated
detection might be necessary to detect abusive groups, privacy advocates
raised serious concerns about their deployment (c.f.,~\cite{BugsPockets}).
Our governance mechanisms will all be user initiated, which makes them less
useful for mitigating abusive groups but still critical in mitigating all
other types of abuse. We believe our results would be compatible with
future automated mechanisms, but leave detailed exploration to future work.

\textbf{\emph{Confidentiality:}} In addition to traditional message
confidentiality, we target \emph{governance confidentiality}. This means
that even in the face of a malicious DS and MS, a group's governance
remains private to the group and the amount of information leaked about the
governance state is minimized. So for example, the DS and MS should not be
able to infer a group's current governance state, including who are the
moderators, what policies are being enforced, and whether a user has been
reported to a community moderator.

Complete governance confidentiality should last until the first report is
made to the MS. At this point, the report may reveal some information about
governance state within the group, e.g., that an abusive moderator exists.
But this revelation should be minimal, meaning that governance state and
actions not involved in a particular report remain undisclosed to the
platform (e.g., who else is a moderator remains hidden). If a community
moderator adds someone to the group, that will be revealed to the platform
(a limitation of the underlying MLS protocol we utilize), but removing
members from the group will not be.

Similarly, and like in other prior works on content-based moderation in
encrypted messaging~\cite{MessengerSecretConvo,SMFCommitting,AMF}, we will
also ensure confidentiality of unreported messages. In our hierarchical
reporting structure, this surfaces as confidentiality from community
moderators and the MS in the case of direct messages (DMs) and from the MS in the case of group messages (GMs).

\textbf{\emph{Further non-goals:}} In addition to our non-goal of automated
detection of fully abusive groups and denial of service, we also do not
focus on resisting traffic analysis attacks. It could be that the pattern
of encrypted messages sent reveals, for example, information about
governance actions, such as who is a moderator. Whether such attacks work
and, if so, what obfuscation strategies can be used is an interesting
question for future work.

We also do not target prevention of bad policies:
their design and the criteria that define what
makes a robust policy are beyond the scope of this work. Poorly or maliciously designed
polices could have loopholes that enable abuse or circumvention, much in
the same way a badly specified multi-party computation functionality can
directly reveal secret inputs. Other work addresses
questions of policy design and specification~\cite{wang2024pika}. As
new insights arise in the literature on policy design, our system will enable rapid prototyping of these ideas in
an encrypted messaging setting.

We also do not specifically target private governance while achieving
metadata privacy, i.e., sender or receiver
anonymity~\cite{sealedsender,Vuvuzela,Riposte,Asynchromix,Tor}. That said,
our layered approach means that if one can build a metadata private
encrypted messaging system that provides an ordered message primitive, then
our techniques would also work in that context. That said, we are not aware
of any metadata-private systems that can easily provide ordered messaging
primitives for groups, and so this remains an open problem.

Finally, we do not target deniability~\cite{OTR} and, relatedly, coercion
resistance~\cite{coercionResistantVoting}. MLS does not provide
deniability. But just like metadata privacy our system should be adaptable
to a deniable encryption layer
(e.g.,~\cite{OTR,DBLP:conf/ccs/RaimondoGK06,DBLP:journals/popets/UngerG18,x3dh}),
by replacing our use of non-deniable digital signatures with deniable
message franking tools~\cite{AMF}. Policies that include voting are subject
to coercion, and again asymmetric message franking would help make the
system more coercion resistant since they would not be as useful as proof
to a coercer of how someone voted.

\section{The Messaging Layer}
\label{sec:extmls}

In this section, we describe how to construct an E2EE messaging layer that
supports our governance layer. Our encrypted messaging layer  will also be
useful more broadly for privacy applications that require synchronized,
distributed state. We start from the messaging layer security (MLS) draft
standard~\cite{mls-protocol,mls-architecture} which provides encrypted
group messaging with strong security properties and scales to large group
sizes.  Then we describe a crucial---but in hindsight
straightforward---modification that endows MLS clients with the ability to
maintain synchronized application state. We will use this capability for
governance, but note that our extension to MLS is likely of broader
interest.

\paragraph{MLS Overview}
MLS relies on an authentication service (AS) for maintaining user
credentials and on a delivery service (DS) for transmitting messages. The
MLS protocol~\cite{mls-protocol} supports efficient encrypted group
messaging with strong guarantees including confidentiality, authenticity,
integrity, forward secrecy, and post-compromise security. We provide more
background on MLS in \apref{sec:mls-background}. Content messages in MLS
are not expected to be broadcast in a consistent order, hence we refer to
them as unordered application messages (UAMs). On the other hand, messages
that update cryptographic state must be consistently ordered.

\paragraph{Consensus in MLS}
MLS uses a simple consensus mechanism for shared cryptographic state:
updates to this state are conveyed through \emph{proposals} and
\emph{commits}. Proposals convey an intention to carry out an action, such
as adding someone to the group. The MLS protocol currently supports eight
proposal types, most of which have to do with group membership. Add,
Remove, and Update are examples of MLS proposal types that allow adding a
group member to the TreeKEM, removing one from it, or allowing a user to
update their public KEM key (e.g., for forward secrecy). A proposal message
contains information relevant to carrying out its associated operation. For
instance, an Add proposal contains the cryptographic information of a user
to be added to the group. By default, proposal messages are private
messages.

A commit message contains one or more proposals, and when a client
processes a commit, they carry out the actions specified in those
proposals, advancing to a new \textit{epoch}, which is the fundamental and
atomic unit of shared group state in MLS. The history of group state
evolution in MLS proceeds in epochs with commit messages referencing the
epochs directly before them. Commit messages are likewise by default
private messages. In our implementation, both proposal and commit messages
are private.

MLS requires that all group members agree on a total ordering over all
commit messages, in order to ensure a linear progression of epochs.
Relatedly, MLS clients must have an established way for dealing with the
scenario in which two or more clients attempt to submit conflicting commits
building off of the same prior epoch. The MLS protocol does not specify a
concrete way for clients to agree on ordering or handle conflicting
commits; these are left as implementation choices for MLS clients and
servers.
For our design, we
assume a strongly consistent DS that provides a consistent message ordering
to clients, which handle conflicts by merging commits that arrive first
according to the server-provided ordering.

\paragraph{Ordered application messages}
A key insight underlying our approach to governance is developing a method
for maintaining a consistent, shared view of governance state within MLS
groups. Looking ahead, governance state will include information on user
roles, group topic, information regarding ongoing governance processes such
as votes. This requires that we extend MLS to allow transmitting arbitrary
application-defined data in a way that is atomically updatable; this is
already solved within MLS to support shared cryptographic material. We
therefore can use the same consensus mechanism (proposals and commits) to
maintain arbitrary shared group state.

We extend MLS to include a new proposal type called an \textit{ordered
    application message (OAM) proposal}. It contains an arbitrary sequence of
bytes, similarly to application messages. The difference is that by
introducing this as a proposal type, we force the sequence of bytes to be
committed to: clients must agree on this string and when it was committed
to. Committing an OAM proposal triggers an epoch change. In addition to
ordered application messages, we devise a mechanism for bootstrapping
existing governance state for new group members and eventual detection of
incorrectly supplied initial state. We call this mechanism group state
announcement and describe it in detail in \secref{sec:design}.

Our design includes a DS that provides a total ordering on commit messages.
When a client sends a commit message to the DS, the DS includes in its
response a list of messages intended for the client that arrived before its
commit message, according to the server's ordering. If this list contains
no other commit messages that attempt to build on the same previous epoch
as this client's commit message, the client proceeds to merge their commit,
applying the proposals contained within it and progressing to a new epoch.

As already mentioned in the MLS specification~\cite[\S14]{mls-protocol},
clients can end up starved of the ability to submit a commit. This means
that not all application features are suitable for implementing via OAM
proposals, e.g., switching regular text messaging to have ordering would
have significant negative impact on performance. However, we do not have
the same requirement of consistency of text message ordering as we do for
governance state updates. Since we expect governance state updates to be
sent with less frequency than user text messages, our usage of OAMs
achieves consistent state evolution with performance. The only case in
which we would expect a high volume of ordered messages in a short period
of time is with voting, and we describe an optimization in
\secref{sec:design} that is able to handle this scenario effectively. We
empirically evaluate the performance of our approach in
\secref{sec:evaluation}.

\paragraph{The messaging layer API} To make precise the interface between
our extended version of MLS and our governance layer, we define a messaging
layer API. This API maps between higher-layer requests (in our context, the
governance layer) to underlying MLS message types. We follow as closely as
possible the OpenMLS API~\cite{openmls}. OpenMLS is a mature implementation
of the MLS protocol. More pragmatically, we will build off OpenMLS in our
implementations. A summary of our messaging layer API is shown in
\figref{fig:mls-api}. Here we focus on the subset of the API related to
messages and group maintenance.
Each API function call maps to one or more underlying MLS protocol
messages types.

\begin{figure}[]
    \center
    \scriptsize
    \begin{tabular}{p{0.55in}p{1.2in}p{.94in}}
        \textbf{Function}             & \textbf{Description}                            & \textbf{MLS messages} \\
        \midrule
        $\sendUam$                    & Send bytes to group, without ordering guarantee &
        Application ~msg                                                                                        \\\midrule $\sendOam$ & Send bytes to group, with
        consistent ordering guarantee & OAM P+C                                                                 \\\midrule $\add$ & Add user to group & Add P+C,
        Welcome~msg                                                                                             \\\midrule $\remove$ & Remove user from group & Remove
        P+C                                                                                                     \\\midrule $\update$ & Update user's public keys & Update
        P+C                                                                                                     \\\midrule $\getepoch$ & Retrieve current epoch number for a group
                                      & N/A                                                                     \\\midrule
    \end{tabular}
    \caption{Functions exposed by our message layer API, including inputs and
        the MLS messages invoked to handle the API request in our implementation.
        Here P+C stands for the indicated proposal type followed immediately by a
        commit.}
    \label{fig:mls-api}
\end{figure}

\section{The Governance Layer}
\label{sec:design}

\begin{figure*}[t]
    \centering
    \begin{tabular}{lcccc}
        \toprule
        \textbf{Action}  & \textbf{Description}                              & \textbf{Content State} & \textbf{Governance State} & \textbf{MLS Messages} \\ \hline
        Text message     & append message to message history                 & \cmark                 &                           & UAM                   \\
        Invite           & update cryptographic state and share with invitee &                        & \cmark                    & OAM, Add, UAM         \\
        Kick             & update cryptographic state and share with kickee  &                        & \cmark                    & OAM, Remove           \\
        Rename Group     & modify group name in shared group state           &                        & \cmark                    & OAM                   \\
        Define Role      & define a new role as a set of allowed actions     &                        & \cmark                    & OAM                   \\
        Assign Role      & assign a role to a user                           &                        & \cmark                    & OAM                   \\
        Content takedown & remove specified content                          & \cmark                 &                           & UAM                   \\
        Report           & send a report of received message                 & \cmark                 &                           & UAM                   \\
        Vote             & send a vote on a proposed action                  &                        & \cmark                    & OAM or UAM            \\
        Accept           & acknowledge acceptance of invitation to group     & \cmark                 &                           & UAM                   \\
        \bottomrule
    \end{tabular}
    \caption{A subset of our supported actions, whether they affect
        governance state or content state, and the MLS messages they produce. UAM refers to an unordered application message, OAM refers to an ordered
        application message, and Add/Remove refers to a commit with the
        Add/Remove proposal. }
    \label{fig:actionsTableFull}

\end{figure*}

The messaging layer described in the last section provides cryptographic
APIs upon which we can build a \emph{governance layer}. This layer sits
between applications and messaging, interposing on application layer
requests to apply policies. A pseudocode specification of our core
governance layer logic can be found in \apref{ap:pseudocode}.

Recall from \secref{sec:setting} our example of a governance flow: (1) a
user~$\user_2$ proposes a vote to change community guidelines; (2)
$\user_3$ harasses $\user_2$ after the guidelines change; (3) $\user_2$
sends an abuse report to community moderator~$\user_1$; and finally (4)
$\user_1$ escalates by reporting the abuse to the platform moderator. In
this section we detail the abstractions that our governance layer provides
to realize such governance actions, and how they are implemented on top of
our messaging layer.

\paragraph{Community structure} Our governance layer organizes users into
groups, which are also tagged with community identifiers. A group consists
of two or more users involved in a shared messaging channel. We do not make
a distinction between DMs and GMs, rather these are both implemented as
groups (of size two or more, respectively). We assume distinct namespaces
for usernames, group identifiers, and community identifiers. These are
assumed public, but users can choose them to be semantically meaningless
(like random numbers) --- we support having internal groups names that
remain private to the current members and can be chosen and modified by
members. A user can initiate a group by invoking a group creation
API call at the messaging layer. Then they can then add users via the
messaging layer. The initiator of a group and group membership is
platform-visible.

We associate to all users a public key and secret key for use by the
governance layer, we refer to these as governance keys. Clients pick their
governance keys, and the public key is registered with the AS like other
public keys used in the messaging layer. As we discuss below, adding a
separate pair of keys allows for clean separation of the layers, while
incurring little performance impact, as we show.

Our governance layer supports a rich role-based access control
(RBAC)~\cite{rbac} mechanism. The RBAC gates who can perform what kinds of
governance actions (a notion we will define soon), what we refer to as a
permission. A set of permissions defines a role. This will support, in our
running example, having a community moderator~$\user_1$ that has the
permissions to remove $\user_3$, while other users do not have this
ability. Our RBAC mechanism is more expressive than recent suggestions for
cryptographic group administration~\cite{cryptoAdmin}.

\paragraph{Governance and content state} To support governance mechanisms,
we need some consistent state replicated across clients. We refer to this
state as the \emph{governance state}. The governance state is a key-value
store that includes information such as user roles and privileges within
the group, current policies like prohibited word lists, and the group name.

We delineate between governance state and all other group-related state,
such as sent plaintext text or image content that's been sent to the group
or via DMs. We refer to this non-governance state as \emph{content state}.
A key enabling architectural decision is that we can separate between
aspects of group state for which governance only works should clients agree
on that state, versus other state for which it is allowable for clients'
views to diverge. This is important for performance and deployability: we
show that useful governance policies can be implemented even when many
aspects of shared group state are potentially inconsistent. In our running
example, the group guidelines are part of the governance state because the
group must agree on the current policies, but individual messages and
reports end up as part of the content state. This means the group does not
``agree'' on the fact that a report occurred, but the moderator $\user_1$
can verify that $\user_3$'s harassment occurred while the group agreed on a
no-swearwords policy.

When a new user joins a group, the inviter includes in the welcome message
a serialization of the current governance state. We assume some efficient
canonical way of encoding the governance state. We use JSON in particular.

\paragraph{Actions} To guide and reason about how community state changes
over time, our governance layer defines a set of possible actions. An
\emph{action} is a message broadcast within a group that can produce
changes to the shared group state or content state. Examples of actions
include sending a text message and changing the group name. We classify
actions, depending on whether they affect the governance state, into two
categories: \emph{governance actions} which can change the governance state
and the content state, and \emph{content actions} which can only change the
content state. For example, sending a text message to a group only changes
the content state, so it is a content action, while changing the group name
changes the governance state, so it is a governance action. We provide more
examples of actions and which types of state they may modify in
\figref{fig:actionsTableFull}. All application-layer requests result in one or more
actions.

To communicate an action, the governance layer prepares an action message.
A header is constructed that includes the sender username, an action ID (a
unique identifier for the action), the group ID, and the community ID.
An unambiguous encoding of the action-related data follows, such as the
plaintext data for sending content or the new group name. Finally, all this is
serialized and signed using the sender's governance digital signing key.
This signature is checked once a message is received, before it is
evaluated by the RBAC or policy engine. While in theory we could utilize
the messaging layer's digital signatures to provide accountability for
action messages, using governance keys allows for clean separation between
abstraction layers --- otherwise we would have to modify the messaging API
to expose low-level details of how MLS messages are framed. It also means
we can swap out our MLS messaging layer with any API-compatible
variant.

As an example of an action, consider when $\user_2$ wants to report
to~$\user_1$ the abusive DMs received from~$\user_3$. To do so, $\user_2$
builds a report action whose payload consists of one or more
action messages. In this case it would be $\user_3$'s DMs, which were,
themselves, action messages that are signed. The report therefore includes
a full serialization of the reported action messages, including their
signatures, making it possible for~$\user_1$ to verify them.

\paragraph{Updating state via OAMs}
Content actions result in a call to $\sendUam$ in the messaging layer,
while governance actions use ordered messages via $\sendOam$. Governance
actions need to use ordered messages to ensure that governance state is
modified consistently across all clients. Since there are no ordering
guarantees on UAMs, using them to encode governance actions can potentially
fork governance state, leading to security and correctness issues. An
ordered application message, when processed by a client, potentially
changes the shared group state. As all clients begin from a common state
and process OAMs in the same order, they retain a consistent group state.

\paragraph{Initializing governance state} While ordered messages allow
current group members to perform consistent governance state updates, we
require a separate mechanism to provide newly invited members with their
initial governance state. In our design, we have a designated action for
group state announcement, which contains an encoding of the group state at
the epoch during which the new member joins. The group state announcement
is sent by the inviter as an UAM. The new member assigns their current group
state to the one in the encoding and includes a hash of this state in their
\texttt{Accept} message, which is broadcast to the entire group. If a
member detects this hash to be inconsistent, they can alert the new member
and group (and/or platform) moderators. We analyze the security of this
approach in \secref{sec:security}.

There are alternate possible approaches in which a group state announcement
could be sent via an ordered message that immediately follows the
\texttt{Add} message. However, doing so results in a more complex state
machine, is worse in terms of performance, and can lead to DOS attacks. In
contrast, our approach has existing group members check the \texttt{Accept}
message of the new member for consistency with their current view of the
governance state.

\paragraph{Policies} We support developer-defined policies. A policy is an
arbitrary process defined using code which determines whether an action
should be executed, in the context of a particular group. Here we adopt the
viewpoint of, and some of the concepts underlying,
PolicyKit~\cite{policykit}, which built powerful governance mechanisms for
settings where all traffic can be seen by a service.
To support privacy we position policy enforcement at the clients, by way of
a policy engine that is applied to broadcast actions. Policies must be
written such that only actions broadcast in OAMs update shared state.
This guarantees governance state consistency.

The policy engine defines an execution model for determining if actions
should proceed. Following PolicyKit, we hardcode that our RBAC engine takes
precedence, so that when processing an action the engine first checks if
the user has a role whose permissions allow the action. If so the action is
invoked. For example, a user who has the moderator role will typically have
the permission to immediate remove a user from the group. If an action
cannot immediately pass according to the RBAC, it is fed into the policy
engine, which will determines when, if at all, the
action will pass.

As with PolicyKit, our policies are defined by a template consisting of
several functions that get called from within the policy engine:
(1) \filterproc defines the scope of the policy. It takes as input an
action message and returns a boolean to indicate whether it is relevant to
the policy. This allows policies to choose which kinds of actions they
impact.
(2) \initproc defines how to initialize state for the policy's execution
of a specific action, and can potentially modify the client state
(3) \checkproc evaluates an action. It takes an action and returns one
of \texttt{passed}, \texttt{failed}, or \texttt{proposed}. The latter
allows for policies that can't yet determine whether they pass or fail,
such as when handling a vote action that requires waiting some period of
time for votes to arrive.
(4) \passproc determines the effect if the action passed. It
takes as input the action and temporary state, and it modifies the
governance or content state.
(5) \failproc determines the response if the action failed.
Normally this routine does nothing, but in some cases policies may want
to display a message to users or clean up state
All functions have access to the governance and content state held by the
client.

The policy engine executes policies as follows. The engine is called
anytime an action is broadcast to the group and does not pass the RBAC. The
engine loops over all policies in a predetermined, developer-defined order.
For each policy, the engine calls in turn \filterproc, \initproc, and then
\checkproc. The first policy for which \filterproc returns \texttt{true}
will be the policy that governs this action. If \checkproc returns
\texttt{proposed}, then the engine keeps track of the action as pending. If
it returns \texttt{passed} then the engine calls \passproc, which executes
the action and otherwise calls \failproc. Once an action passes or fails,
these routines are expected to clean up state allocated by \initproc. The
policy engine will periodically attempt to pass proposed actions by
re-running \checkproc, whose output can change based on the policy's
current state.

\paragraph{Voting} Although our policy framework is quite general, we are
interested in how it can enable collective action among users to effect
change within the community. A classic example of this is voting, in which
community members indicate their preference for a change, and carry it out
if there is sufficient support. Our policy framework allows the expression
voting procedures for arbitrary actions, including changing the group name
and promoting users to moderator status.
Individual votes can be cast within ordered messages, however, our design
admits an optimization for high-volume voting behavior. If many votes were
to be cast simultaneously through individual ordered messages, there would
be a high degree of contention and many retransmissions of votes. However,
votes for a single poll can be aggregated in any order if we consider
simple majority or threshold votes. Therefore, we allow clients to send
votes in unordered messages. Multiple votes can then be batched into a
single commit (for instance, when enough messages are collected to bring a
vote to completion), at which point they are fed into the policy engine. As
a result, a group can process many votes within a short period of time, as
we show in \secref{sec:evaluation}.

\paragraph{Hierarchical governance}
\label{ssec:platform-level-moderation}
All the above facilities support community governance in a way that is, by
design, opaque to the platform. But we also want to allow community members
to escalate problems to moderators run by the platform, such as in our
example when $\user_1$ reports $\user_3$ to the platform for their
behavior. We refer to this as hierarchical governance.

Our governance layer therefore includes the ability to interact with a
\emph{moderation service (MS)} operated by the platform.  This service can
have the ability to receive reports, process reports, and limit users at
the platform level, e.g., by instructing the AS to revoke a user's keys or
temporarily blocking them from sending messages.

While there are multiple ways to build an MS, we do so by setting aside a
distinguished username, e.g., \texttt{@moderation}, which is authorized for
taking platform moderation action and receiving reports on behalf of the
platform. This design choice enables us to reuse existing infrastructure
and allow for conversations surrounding reported content. We define a
structured report as a plaintext byte string consisting of a serialization
of a username, sequence of one or more action messages, and an (optional)
reason for the report (an arbitrary byte string in our current
implementation). The structured report is sent to \texttt{\@moderation} via
$\sendUam$ in a group that consists of the user and \texttt{\@moderation}
(the group has some distinguished group identifier).

As implied by using a UAM, we do not require consistency: delayed or
dropped reports can be resent by the user just like regular messages. The
moderation service can verify digital signatures in the included action
messages, providing reporter accountability.
Unlike in traditional platforms, the technical capabilities of the MS are
limited to user-level limiting (since communities are implemented
client-side). In particular, the moderation service can block a user at the
platform-level (either indefinitely or for a limited time), but cannot
block specific messages, groups, or communities based on their content.

\section{Security Evaluation}
\label{sec:security}

\newcommand{\myemph}[1]{\textbf{\emph{#1}}}

In this section, we analyze \sysname in terms of its ability to achieve our
security goals: governance integrity, accountability, and confidentiality.

Prior work has analyzed the MLS
protocol~\cite{modDesignMsg,insiderSecMLS,treeSync}, establishing that it
achieves strong message confidentiality and authenticity. MLS additionally
provides post-compromise and forward secrecy, though we will not need this
for our subsequent analyses.
Note that the modifications we made to the MLS messaging layer (ordered
application messages) reuse the existing underlying proposal plus
commitment mechanisms, and therefore inherits their security.

To analyze governance extensions with respect to our security definitions
(\secref{sec:setting}), we enumerate possible attacks. For each attack, we
analyzed the extent to which our system prevents, mitigates, or allows the
eventual detection of the attack. This is in line with the methodology used
in the Tor paper~\cite{Tor} and in the MLS Architecture Specification
\cite{mls-architecture} to establish confidence in the security of complex
protocols. Additional discussion of
out-of-scope attacks and security goals appears in
\apref{sec:add-security}.

\paragraph{Attacks against integrity} We analyzed attacks that seek to
undermine a group's governance state. At a high level, the authenticity of
MLS and its hash transcript for epochs ensures that honest clients will
reject maliciously generated state updates. Even with a colluding,
malicious DS, the best an adversary can do in most cases is partition the
group forever, a form of denial of service since it means the partitioned
honest users can never be allowed to talk to each other again (lest they
detect the attack). In more detail, we considered attacks including policy
violation, impersonation, governance state partitioning, and invalid
initial state, and vote suppression:

\myemph{Policy violation:} Suppose malicious clients collude to attempt to
violate the policy of the group by trying to perform an unauthorized
action, such as one malicious client performing an action outside that
client's RBAC-defined role. Since all honest clients have the same
governance state and run the same code to interpret the governance state,
they will not accept this unauthorized action. This is true even if the
malicious clients collude with a malicious DS.

\myemph{Impersonation:} A malicious client or DS could attempt to
impersonate a member of the group and send messages as that other member.
But the authenticity of MLS messages plus our assumption that the AS is
trustworthy rules out such impersonating messages being accepted by honest
clients.

\myemph{Governance state partitioning:} Suppose a malicious client colludes
with a malicious DS in an attempt to produce inconsistent governance state
within the group. This means that the goal is to have two distinct subsets
$A,B$ of the group have different governance states; $A$ and $B$ must have
at least one honest client each. We refer to this as a partitioning attack
on the governance state. Since the governance state can only be updated by
commit messages that are included in the MLS hash transcript, such a
partitioning attack can proceed only as long as no honest client in $A$
receives a message from $B$ (or vice versa), as the first such message will
not verify by the recipient. Thus, the adversary can at best split the
group and never allow future communication.

\myemph{Invalid initial state:} Any malicious client, regardless of their
RBAC permissions, can send an Invite message with an incorrect initial
governance state. For example, consider our running example group, we could
have that the abusive user $U_3$ instead behaves maliciously and invites a
new member $U_4$ to the group, but providing an initial governance state
that does not include the policy against swearing and has $U_3$ with the
RBAC role to add users.
Honest clients will reject this invite message, but the newly invited
client does not know that this is invalid. However, when the new member
sends an \texttt{Accept} message, this message will contain a hash of the
received governance state. By collision resistance, that hash will not
match the one expected by honest clients. Thus, while $U_4$ may send
additional messages or interact with $U_3$, no honest user will accept
$U_4$'s messages and, moreover, as soon as they come online, they will
detect that an attack occurred. Thus, even with collusion by the DS,
the malicious client can at best partition the group.

\myemph{Vote suppression:} A malicious DS may attempt to prevent one or
more client's votes from counting. Because votes are encrypted, it isn't
directly revealed to the DS which are votes (and for what election). Thus,
naively the DS would have to just drop all messages emanating from the
target clients. But even if the DS can somehow precisely target UAMs and
OAMs for dropping, a client can still detect if their votes are being
suppressed: the transcript hash consistency mechanism enables clients to
obtain a consistent ordering over commits, which contain all the registered
client votes.

\paragraph{Attacks against confidentiality} Recall that for confidentiality
we want to, by default, ensure the privacy of group content and governance
actions from a malicious DS. This covers other confidentiality threats,
like network adversaries, who see less than the DS. Here we rely primarily
on the confidentiality of MLS messages, and do not consider traffic
analysis attacks here (see discussion at the end of this section). We
considered the following attacks:

\myemph{Inferring the content of governance state:} A malicious DS may
attempt to infer the content of the governance state based on the
transcript of messages it relays on behalf of clients. For example, the DS
might want to identify moderators or admins. But all updates to governance
state are sent through encrypted commit messages. Group state announcements
that supply new members with the current governance state are sent via
encrypted UAMs. By the confidentiality of the encryption scheme MLS uses,
the DS cannot observe the content of the governance state.

\myemph{Inferring content of unreported messages:} All messages and
reporting signatures are encrypted via MLS, and MLS' confidentiality
guarantees ensure that even learning about one message does not leak any
additional information about another encrypted plaintext. As a result, even
a malicious DS would not be able to infer anything about the contents of
unreported messages  beyond what is revealed by a reported message.

\myemph{Inferring content of user votes:} An adversarial DS may attempt to
learn the value of votes that clients cast for policies that involve
voting. Vote messages are encrypted through MLS, and therefore remain
confidential as the DS observes only ciphertexts.

\myemph{Inferring outcome of a vote:} The DS could attempt to learn the
outcome of a vote. However, votes are encrypted and aggregated on client
devices. After aggregation, the change a vote produces, if successful, is
applied to the local governance state. As a result, the DS cannot infer the
result of a vote.

\paragraph{Attacks against accountability} A malicious user may attempt to
circumvent accountability either by sending a message that is accepted by a
recipient but unreportable to a moderator or framing an honest sender for
having sent an incriminating message. We prevent both types of attacks via
the authenticity properties of digital signatures used at the governance
layer. Recall that deniability is not a goal of our system since MLS itself
does not provide it.

\myemph{Report evasion:} A malicious user may attempt to arrange for their
messages to not be reportable, violating what is often called sender
binding. An attack here seeks to send a message that verifies for an honest
recipient, but does not verify for a moderator. Since we use a standard
digital signature, and we trust the AS to provide correct signature public
keys, these verification procedures are equivalent, ruling out such sender
binding attacks. We note that this also covers moderators and other
privileged users within the group, and that all UAMs and OAMs are
reportable, including those associated to actions.

\myemph{Fake reports:} A malicious user can attempt to frame a user,
violating what is often called receiver binding. Here the malicious client
tries to trick a moderator into accepting a reported message that was not
sent by the honest user specified in the report.  Because we trust the AS,
doing so would result in an existential forgery against the digital
signature scheme.

\section{Implementation and Evaluation}
\label{sec:evaluation}

\begin{figure*}[t]
    \centering
    \setlength\tabcolsep{4pt} %
    \footnotesize %
    \setlength{\belowcaptionskip}{-10pt}
    \newcolumntype{d}[1]{D{.}{.}{#1}} %
    \newcolumntype{L}[1]{>{\raggedright\arraybackslash}p{#1}}
    \begin{tabular}{L{3cm} D{.}{.}{3} D{.}{.}{3} D{.}{.}{3} D{.}{.}{3} D{.}{.}{3} | D{.}{.}{3} D{.}{.}{3} D{.}{.}{3} D{.}{.}{3} }
        \toprule
                                            & \multicolumn{5}{c|}{\textbf{Operation Latency \& Traffic}} & \multicolumn{4}{c}{\textbf{Sync$^s$ Latency \& Traffic}}                                                                                                                                                                                                                                                                                          \\
        \addlinespace
        \multicolumn{1}{c}{\textbf{Action}} & \multicolumn{1}{c}{\textbf{Request}}                       & \multicolumn{1}{c}{\textbf{Network}}                     & \multicolumn{1}{c}{\textbf{Post-}}      & \multicolumn{1}{c}{\textbf{Total }}  & \multicolumn{1}{c|}{\textbf{Traffic}} & \multicolumn{1}{c}{\textbf{Network}}  & \multicolumn{1}{c}{\textbf{Message}}    & \multicolumn{1}{c}{\textbf{Total }}  & \multicolumn{1}{c}{\textbf{Traffic}}
        \\
                                            & \multicolumn{1}{c}{\textbf{Gen.}}                          & \multicolumn{1}{c}{\textbf{Overhead}}                    & \multicolumn{1}{c}{\textbf{Processing}} & \multicolumn{1}{c}{\textbf{Latency}} &                                       & \multicolumn{1}{c}{\textbf{Overhead}} & \multicolumn{1}{c}{\textbf{Processing}} & \multicolumn{1}{c}{\textbf{Latency}} &                                      \\
                                            & \multicolumn{1}{c}{(ms)}                                   & \multicolumn{1}{c}{(ms)}                                 & \multicolumn{1}{c}{(ms)}                & \multicolumn{1}{c}{(ms)}             & \multicolumn{1}{c|}{(KB)}             & \multicolumn{1}{c}{(ms)}              & \multicolumn{1}{c}{(ms)}                & \multicolumn{1}{c}{(ms)}             & \multicolumn{1}{c}{(KB)}             \\
        \midrule
        Invite (for 63 invitees)            & 4.78                                                       & 650.96                                                   & 12.62                                   & 823.47                               & 636.94                                &                                       &                                         &                                                                             \\
        Add (for 63 invitees)               & 20.76                                                      & 360.39                                                   & 0.15                                    & 485.25                               & 176.65                                & 482.88                                & 2.03                                    & 486.18                               & 124.92                               \\
        GovStateAnn.                        & 0.58                                                       & 258.08                                                   & 0.12                                    & 259.02                               & 9.62                                  &                                       &                                         &                                      &                                      \\
        \midrule
        Accept$^s$                          & 1.28                                                       & 254.2                                                    & 0.02                                    & 255.72                               & 3.65                                  & 625.19                                & 28.77                                   & 655.29                               & 223.79                               \\
        \midrule
        RenameGroup                         & 10.18                                                      & 350.0                                                    & 0.13                                    & 360.68                               & 44.91                                 & 404.32                                & 1.88                                    & 407.35                               & 56.7                                 \\
        \midrule
        VotePropose (Rename)                & 10.12                                                      & 313.45                                                   & 0.16                                    & 324.07                               & 33.95                                 & 392.95                                & 1.87                                    & 395.99                               & 51.22                                \\
        \midrule
        Vote$^s$ (Rename)                   & 0.37                                                       & 255.58                                                   & 0.05                                    & 256.25                               & 3.89                                  & 720.08                                & 37.65                                   & 759.04                               & 378.59                               \\
        \midrule
        Send (10-char Text)                 & 0.40                                                       & 257.13                                                   & 0.08                                    & 257.87                               & 3.54                                  & 371.85                                & 0.63                                    & 373.71                               & 37.32                                \\
        \midrule
        Send (100-char Text)                & 0.40                                                       & 257.4                                                    & 0.08                                    & 258.13                               & 3.86                                  & 360.62                                & 0.60                                    & 362.38                               & 37.65                                \\

        \bottomrule
    \end{tabular}
    \caption{Latency breakdown for various messaging operations and traffic for a user in a group of 64 users in \sysname (5-trial average). Clients are on US-East AWS instances while DS/AS are on US-West. For operations$^s$ requested by multiple clients simultaneously, data from the 33rd starting client is used. Total latency represents the time between loading the pre-operation group state and saving the post-operation group state, including server processing and key package generation/update delays. Sync request generations take consistently $<0.01$ ms. For results in a group of 1024, refer to \figref{fig:netTimes1024} in \apref{addl_exp}.}

    \label{fig:netTimes}
\end{figure*}

We now describe a concrete realization of our governance approach: a
proof-of-concept messaging platform, called \sysname, that supports an
expressive set of governance policies. This implementation helped us
explore the ease with which developers might build platforms with rich
governance features.
It also allowed us to assess performance overheads
relative to governance-free encrypted messaging.

\subsection{The Platform: \sysname}

We call our platform prototype \sysname. We forked the Rust implementation
OpenMLS \cite{openmls-link} to add our new ordered
application message proposal type and to modify the exposed API as needed.
The changes to the library are minimal, reflecting our goal of modular design.
We then implemented our governance layer logic in Rust. It totals \locGovernanceLayer{}
lines of code as counted by the \texttt{cloc} utility, not counting
specific policies.

We implemented \sysname by designing a set of policies, as we elaborate on
below. We constructed a CLI client in Rust, totaling \locClientCode{} lines of code. It is capable of managing histories and states for multiple groups in multiple communities. Additionally, we developed simple yet efficient DS and AS services, also in Rust.

\paragraph{Delivery service architecture} Our delivery service is
responsible for ferrying ordered, unordered, and welcome messages between
clients. Additionally, the DS distributes user-submitted cryptographic material
(KeyPackages), which are used by other users to add them to MLS groups. In line
with our asynchronous setting, users send and receive messages via issuing
requests. There are separate requests for handling ordered and unordered
messages. In our implementation, ordered messages are inserted into per-group synchronized queues to ensure total ordering.
Unordered messages are placed in individual per-user queues. This means that we
have that group membership is revealed to the DS (recall that we do not target
membership privacy). When users send ordered messages, in the response, they
receive all ordered messages that arrived before theirs in that group. This enables clients
to break ties among ordered messages to ensure consistency, all while ensuring minimal lock usage
which is important for achieving high throughput.

\paragraph{Included governance features}
\sysname includes RBAC that enables flexible permissions hierarchies among
group members. For instance, our system can support having admins that can
remove users from a group, add new members to the group, takedown content,
change the (private) group name, assign a new moderator, and more.
Importantly, admins have the ability to add more RBAC roles and
permissions; a common pattern we expect is to have admins delegate to
moderators the ability to takedown content and remove regular users (except
the admins). But this is just an example that we implemented, and different
moderation hierarchies are configurable via governance actions.

We also built a policy to support voting.
Any user can initiate a poll to decide on whether to perform a governance
action, which votes to elect new moderators, change the name of the group,
modify community guidelines, takedown some content, or perform other
governance actions. Our initial implementation waits for all current
members of the group to vote, and then executes the governance action, a
rename action in our proof of concept, if a simple majority voted to do so.
It would be easy to modify the policy to instead have the vote end after a
set duration, and take a decision based on who participated (and even set
thresholds on how many need to have participated). As described in
\secref{sec:design}, we optimized the process to minimize the usage of
ordered messages, significantly reducing contention and hence total
latency.
We also support polls that perform no governance actions, as may often be
the case when users want to vote on something that happens off the
platform.

Our experience writing policy code indicates that it enables rapidly
building rich governance features. We have also begun prototyping reputation
systems (modifying user permissions dynamically based on reputation score),
setting word filters to automatically block content (which would give a
mechanism to enforce the profanity ban mentioned in our example from
\secref{sec:setting}), and more. %

\subsection{Performance Evaluation}

\sysname is the first system we are aware of that builds rich extensible
governance features in an E2EE setting. In this section we evaluate the
performance overheads of our governance approach over the baseline of a
basic messaging platform. For the latter we use as baseline a version
\sysname with all governance features turned off. We refer to this as
\baseline. This means that no authorization checks occur, reporting
signatures are not included with messages, and the policy engine is not
run. Our evaluation focuses on assessing the latency and bandwidth
overheads incurred by our approach to governance, as well as its effect on
scalability in terms of group size and server resources consumed.

\begin{figure*}[t]
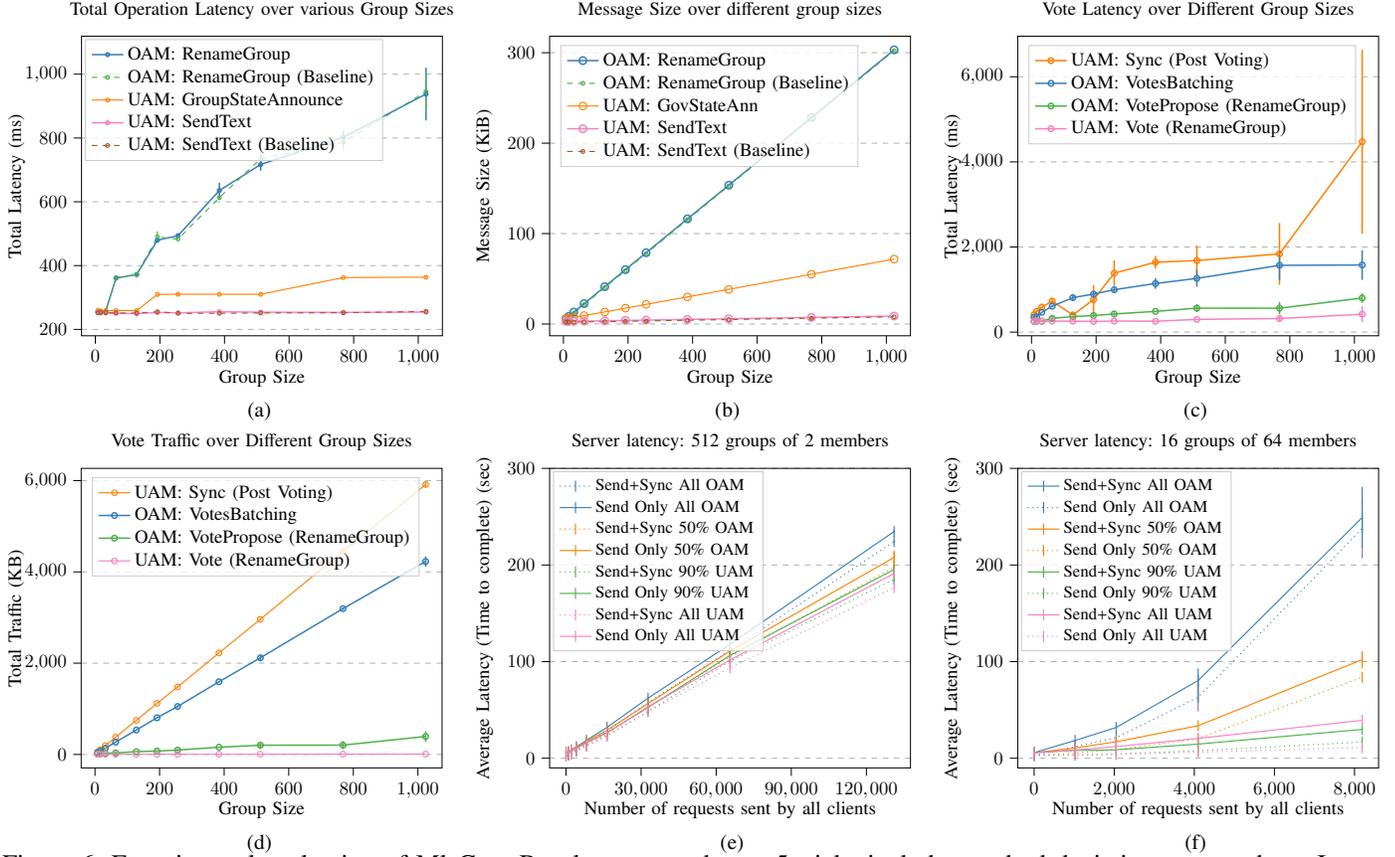

    \centering
    \begin{subfigure}[b]{.3\linewidth}
        \captionsetup[subfigure]{justification=centering}
        \centering
        \input{tikz_micro_time}
        \footnotesize
        \vspace*{-1mm}
        \hspace*{16mm}(a)
        \vspace*{2mm}
        \phantomcaption
        \label{fig:micro-latency}
    \end{subfigure}
    \hfill
    \begin{subfigure}[b]{.3\linewidth}
        \centering
        \begin{tikzpicture}[scale=0.7]

\definecolor{darkgray176}{RGB}{176,176,176}
\definecolor{darkorange2551270}{RGB}{255,127,0}
\definecolor{hotpink247129191}{RGB}{247,129,191}
\definecolor{lightgray204}{RGB}{204,204,204}
\definecolor{mediumseagreen7717574}{RGB}{77,175,74}
\definecolor{sienna1668640}{RGB}{166,86,40}
\definecolor{steelblue55126184}{RGB}{55,126,184}

\begin{axis}[
legend cell align={left},
legend style={
  fill opacity=0.8,
  draw opacity=1,
  text opacity=1,
  at={(0.03,0.97)},
  anchor=north west,
  draw=lightgray204
},
tick align=outside,
tick pos=left,
title={Message Size over different group sizes},
x grid style={darkgray176},
xlabel={Group Size},
xmin=-42.8, xmax=1074.8,
xtick style={color=black},
y grid style={darkgray176},
ylabel={Message Size (KiB)},
ymin=-13.0033284176416, ymax=318.26503688274,
ytick style={color=black},
            ymajorgrids=true, grid style=dashed,
]
\path [draw=steelblue55126184, thick]
(axis cs:8,5.86075073359045)
--(axis cs:8,5.92323364140955);

\path [draw=steelblue55126184, thick]
(axis cs:16,8.26041001119306)
--(axis cs:16,8.34193373880694);

\path [draw=steelblue55126184, thick]
(axis cs:32,13.1043227503899)
--(axis cs:32,13.1308334996101);

\path [draw=steelblue55126184, thick]
(axis cs:64,22.5009476471334)
--(axis cs:64,22.5431929778666);

\path [draw=steelblue55126184, thick]
(axis cs:128,41.2116596870792)
--(axis cs:128,41.3344340629208);

\path [draw=steelblue55126184, thick]
(axis cs:192,60.0257727424201)
--(axis cs:192,60.2320397575799);

\path [draw=steelblue55126184, thick]
(axis cs:256,78.7532974425819)
--(axis cs:256,78.9451400574181);

\path [draw=steelblue55126184, thick]
(axis cs:384,116.219703854814)
--(axis cs:384,116.387718020186);

\path [draw=steelblue55126184, thick]
(axis cs:512,153.395350445029)
--(axis cs:512,153.612462054971);

\path [draw=steelblue55126184, thick]
(axis cs:768,228.168317252491)
--(axis cs:768,228.444573372509);

\path [draw=steelblue55126184, thick]
(axis cs:1024,303.04261608546)
--(axis cs:1024,303.20738391454);

\path [draw=mediumseagreen7717574, semithick]
(axis cs:8,4.88619599206649)
--(axis cs:8,4.93685088293351);

\path [draw=mediumseagreen7717574, semithick]
(axis cs:16,7.32476105968633)
--(axis cs:16,7.37289519031367);

\path [draw=mediumseagreen7717574, semithick]
(axis cs:32,12.1103194841122)
--(axis cs:32,12.1662430158878);

\path [draw=mediumseagreen7717574, semithick]
(axis cs:64,21.5619746709514)
--(axis cs:64,21.6091190790486);

\path [draw=mediumseagreen7717574, semithick]
(axis cs:128,40.2701228927907)
--(axis cs:128,40.3837833572094);

\path [draw=mediumseagreen7717574, semithick]
(axis cs:192,59.0226175469717)
--(axis cs:192,59.1754293280283);

\path [draw=mediumseagreen7717574, semithick]
(axis cs:256,77.7501064737795)
--(axis cs:256,77.9069247762205);

\path [draw=mediumseagreen7717574, semithick]
(axis cs:384,115.141300268855)
--(axis cs:384,115.387605981145);

\path [draw=mediumseagreen7717574, semithick]
(axis cs:512,152.602699485115)
--(axis cs:512,152.721909889885);

\path [draw=mediumseagreen7717574, semithick]
(axis cs:768,227.328640857061)
--(axis cs:768,227.514327892939);

\path [draw=mediumseagreen7717574, semithick]
(axis cs:1024,302.051786444168)
--(axis cs:1024,302.388838555832);

\path [draw=darkorange2551270, semithick]
(axis cs:8,5.90776164179774)
--(axis cs:8,5.95356648320226);

\path [draw=darkorange2551270, semithick]
(axis cs:16,6.37998121999638)
--(axis cs:16,6.39267503000362);

\path [draw=darkorange2551270, semithick]
(axis cs:32,7.33815062326288)
--(axis cs:32,7.38880250173712);

\path [draw=darkorange2551270, semithick]
(axis cs:64,9.31770641871191)
--(axis cs:64,9.33541858128809);

\path [draw=darkorange2551270, semithick]
(axis cs:128,13.2909261707463)
--(axis cs:128,13.3664957042537);

\path [draw=darkorange2551270, semithick]
(axis cs:192,17.4451208790933)
--(axis cs:192,17.5587853709067);

\path [draw=darkorange2551270, semithick]
(axis cs:256,21.6232070707085)
--(axis cs:256,21.7318710542915);

\path [draw=darkorange2551270, semithick]
(axis cs:384,29.906859269939)
--(axis cs:384,30.003296980061);

\path [draw=darkorange2551270, semithick]
(axis cs:512,38.2739992307506)
--(axis cs:512,38.3228757692494);

\path [draw=darkorange2551270, semithick]
(axis cs:768,54.9667917354067)
--(axis cs:768,55.0929738895933);

\path [draw=darkorange2551270, semithick]
(axis cs:1024,71.6934888185801)
--(axis cs:1024,71.8209643064199);

\path [draw=hotpink247129191, thick]
(axis cs:8,2.92273955067924)
--(axis cs:8,2.95929169932076);

\path [draw=hotpink247129191, thick]
(axis cs:16,2.9445566571504)
--(axis cs:16,2.9835683428496);

\path [draw=hotpink247129191, thick]
(axis cs:32,3.04060595644393)
--(axis cs:32,3.07267529355607);

\path [draw=hotpink247129191, thick]
(axis cs:64,3.21109431390504)
--(axis cs:64,3.24085881109496);

\path [draw=hotpink247129191, thick]
(axis cs:128,3.53408231027014)
--(axis cs:128,3.56826143972987);

\path [draw=hotpink247129191, thick]
(axis cs:192,3.90888925913247)
--(axis cs:192,3.94267324086753);

\path [draw=hotpink247129191, thick]
(axis cs:256,4.280078125)
--(axis cs:256,4.320703125);

\path [draw=hotpink247129191, thick]
(axis cs:384,5.03231052567896)
--(axis cs:384,5.05948634932104);

\path [draw=hotpink247129191, thick]
(axis cs:512,5.77825939608964)
--(axis cs:512,5.80963122891036);

\path [draw=hotpink247129191, thick]
(axis cs:768,7.28246378607777)
--(axis cs:768,7.31245808892223);

\path [draw=hotpink247129191, thick]
(axis cs:1024,8.77716419363334)
--(axis cs:1024,8.82557018136666);

\path [draw=sienna1668640, semithick]
(axis cs:8,2.0543245505575)
--(axis cs:8,2.0749723244425);

\path [draw=sienna1668640, semithick]
(axis cs:16,2.08066198494254)
--(axis cs:16,2.09902551505746);

\path [draw=sienna1668640, semithick]
(axis cs:32,2.14959789595081)
--(axis cs:32,2.16290210404919);

\path [draw=sienna1668640, semithick]
(axis cs:64,2.31688867328058)
--(axis cs:64,2.34131445171942);

\path [draw=sienna1668640, semithick]
(axis cs:128,2.64875531940719)
--(axis cs:128,2.68288530559281);

\path [draw=sienna1668640, semithick]
(axis cs:192,3.03212929688125)
--(axis cs:192,3.05654257811875);

\path [draw=sienna1668640, semithick]
(axis cs:256,3.40991554044616)
--(axis cs:256,3.42172508455384);

\path [draw=sienna1668640, semithick]
(axis cs:384,4.15843257300572)
--(axis cs:384,4.17516117699428);

\path [draw=sienna1668640, semithick]
(axis cs:512,4.91382302524006)
--(axis cs:512,4.92250509975994);

\path [draw=sienna1668640, semithick]
(axis cs:768,6.39968137623523)
--(axis cs:768,6.42609987376477);

\path [draw=sienna1668640, semithick]
(axis cs:1024,7.9262243165679)
--(axis cs:1024,7.9479944334321);

\addplot [thick, steelblue55126184, mark=o, mark size=2, mark options={solid}]
table {%
8 5.8919921875
16 8.301171875
32 13.117578125
64 22.5220703125
128 41.273046875
192 60.12890625
256 78.84921875
384 116.3037109375
512 153.50390625
768 228.3064453125
1024 303.125
};

\addplot [semithick, mediumseagreen7717574, dashed, mark=o, mark size=1, mark options={solid}]
table {%
8 4.9115234375
16 7.348828125
32 12.13828125
64 21.585546875
128 40.326953125
192 59.0990234375
256 77.828515625
384 115.264453125
512 152.6623046875
768 227.421484375
1024 302.2203125
};

\addplot [semithick, darkorange2551270, mark=o, mark size=2, mark options={solid}]
table {%
8 5.9306640625
16 6.386328125
32 7.3634765625
64 9.3265625
128 13.3287109375
192 17.501953125
256 21.6775390625
384 29.955078125
512 38.2984375
768 55.0298828125
1024 71.7572265625
};

\addplot [thick, hotpink247129191, mark=o, mark size=2, mark options={solid}]
table {%
8 2.941015625
16 2.9640625
32 3.056640625
64 3.2259765625
128 3.551171875
192 3.92578125
256 4.300390625
384 5.0458984375
512 5.7939453125
768 7.2974609375
1024 8.8013671875
};

\addplot [semithick, sienna1668640, dashed, mark=o, mark size=1, mark options={solid}]
table {%
8 2.0646484375
16 2.08984375
32 2.15625
64 2.3291015625
128 2.6658203125
192 3.0443359375
256 3.4158203125
384 4.166796875
512 4.9181640625
768 6.412890625
1024 7.937109375
};
        \addlegendentry{OAM: RenameGroup}
        \addlegendentry{OAM: RenameGroup (Baseline)}
        \addlegendentry{UAM: GovStateAnn}
        \addlegendentry{UAM: SendText}
        \addlegendentry{UAM: SendText (Baseline)}
    \end{axis}

\end{tikzpicture}
        \footnotesize
        \vspace*{-1mm}
        \hspace*{16mm}(b)
        \vspace*{2mm}
        \phantomcaption
        \label{fig:micro-bandwidth}
    \end{subfigure}
    \hfill
    \begin{subfigure}[b]{.3\linewidth}
        \centering
        \begin{tikzpicture}[scale=0.7]

    \definecolor{darkgray176}{RGB}{176,176,176}
    \definecolor{darkorange2551270}{RGB}{255,127,0}
    \definecolor{hotpink247129191}{RGB}{247,129,191}
    \definecolor{lightgray204}{RGB}{204,204,204}
    \definecolor{mediumseagreen7717574}{RGB}{77,175,74}
    \definecolor{steelblue55126184}{RGB}{55,126,184}

    \begin{axis}[
            legend cell align={left},
            legend style={
                    fill opacity=0.8,
                    draw opacity=1,
                    text opacity=1,
                    at={(0.03,0.97)},
                    anchor=north west,
                    draw=lightgray204
                },
            tick align=outside,
            tick pos=left,
            title={Vote Latency over Different Group Sizes},
            x grid style={darkgray176},
            xlabel={Group Size},
            xmin=-42.8, xmax=1074.8,
            xtick style={color=black},
            y grid style={darkgray176},
            ylabel={Total Latency (ms)},
            ymin=-85.3544292580335, ymax=6952.61046380395,
            ytick style={color=black},
            ymajorgrids=true, grid style=dashed,
        ]
        \path [draw=darkorange2551270, thick]
        (axis cs:8,405.113169173501)
        --(axis cs:8,412.816698426499);

        \path [draw=darkorange2551270, thick]
        (axis cs:16,460.266173347497)
        --(axis cs:16,521.175355852503);

        \path [draw=darkorange2551270, thick]
        (axis cs:32,557.414147012515)
        --(axis cs:32,619.105394587486);

        \path [draw=darkorange2551270, thick]
        (axis cs:64,681.514328346007)
        --(axis cs:64,778.148008453993);

        \path [draw=darkorange2551270, thick]
        (axis cs:128,400.510757395526)
        --(axis cs:128,406.924874104474);

        \path [draw=darkorange2551270, thick]
        (axis cs:192,420.71625828205)
        --(axis cs:192,1106.22615471795);

        \path [draw=darkorange2551270, thick]
        (axis cs:256,1085.16849988037)
        --(axis cs:256,1686.47124771963);

        \path [draw=darkorange2551270, thick]
        (axis cs:384,1494.59088237308)
        --(axis cs:384,1789.70162162692);

        \path [draw=darkorange2551270, thick]
        (axis cs:512,1329.15910355867)
        --(axis cs:512,2034.25566004133);

        \path [draw=darkorange2551270, thick]
        (axis cs:768,1111.54741983901)
        --(axis cs:768,2562.79464766099);

        \path [draw=darkorange2551270, thick]
        (axis cs:1024,2309.82064433524)
        --(axis cs:1024,6632.70296866476);

        \path [draw=steelblue55126184, thick]
        (axis cs:8,346.955101994329)
        --(axis cs:8,354.975720005671);

        \path [draw=steelblue55126184, thick]
        (axis cs:16,352.944337067373)
        --(axis cs:16,359.320364132627);

        \path [draw=steelblue55126184, thick]
        (axis cs:32,462.257359461263)
        --(axis cs:32,471.069715738738);

        \path [draw=steelblue55126184, thick]
        (axis cs:64,589.956023912)
        --(axis cs:64,631.236512488);

        \path [draw=steelblue55126184, thick]
        (axis cs:128,736.783862873274)
        --(axis cs:128,883.833244326726);

        \path [draw=steelblue55126184, thick]
        (axis cs:192,821.612099873094)
        --(axis cs:192,966.755494526906);

        \path [draw=steelblue55126184, thick]
        (axis cs:256,943.796630457382)
        --(axis cs:256,1049.40017634262);

        \path [draw=steelblue55126184, thick]
        (axis cs:384,1016.89738567238)
        --(axis cs:384,1268.25611992762);

        \path [draw=steelblue55126184, thick]
        (axis cs:512,1064.12246081747)
        --(axis cs:512,1461.88924798253);

        \path [draw=steelblue55126184, thick]
        (axis cs:768,1219.14146216627)
        --(axis cs:768,1919.53940783373);

        \path [draw=steelblue55126184, thick]
        (axis cs:1024,1228.95851444122)
        --(axis cs:1024,1926.42193435877);

        \path [draw=mediumseagreen7717574, thick]
        (axis cs:8,250.384209582888)
        --(axis cs:8,255.493255217112);

        \path [draw=mediumseagreen7717574, thick]
        (axis cs:16,250.499517955653)
        --(axis cs:16,255.552985244347);

        \path [draw=mediumseagreen7717574, thick]
        (axis cs:32,253.385773969899)
        --(axis cs:32,260.132473230101);

        \path [draw=mediumseagreen7717574, thick]
        (axis cs:64,271.480781223262)
        --(axis cs:64,376.658329176738);

        \path [draw=mediumseagreen7717574, thick]
        (axis cs:128,361.385215836585)
        --(axis cs:128,366.554487363415);

        \path [draw=mediumseagreen7717574, thick]
        (axis cs:192,348.381056556156)
        --(axis cs:192,431.258173043844);

        \path [draw=mediumseagreen7717574, thick]
        (axis cs:256,391.669067998605)
        --(axis cs:256,461.636152001395);

        \path [draw=mediumseagreen7717574, thick]
        (axis cs:384,483.561247640112)
        --(axis cs:384,491.205972759888);

        \path [draw=mediumseagreen7717574, thick]
        (axis cs:512,469.583476911324)
        --(axis cs:512,656.688393088676);

        \path [draw=mediumseagreen7717574, thick]
        (axis cs:768,420.703774433051)
        --(axis cs:768,705.018768366949);

        \path [draw=mediumseagreen7717574, thick]
        (axis cs:1024,686.799127559444)
        --(axis cs:1024,916.076925640556);

        \path [draw=hotpink247129191, thick]
        (axis cs:8,248.043147927665)
        --(axis cs:8,259.616018872335);

        \path [draw=hotpink247129191, thick]
        (axis cs:16,244.154065494805)
        --(axis cs:16,271.612402105195);

        \path [draw=hotpink247129191, thick]
        (axis cs:32,247.19121598156)
        --(axis cs:32,259.92086481844);

        \path [draw=hotpink247129191, thick]
        (axis cs:64,250.031841938113)
        --(axis cs:64,267.613959261887);

        \path [draw=hotpink247129191, thick]
        (axis cs:128,252.626703044035)
        --(axis cs:128,263.121225355964);

        \path [draw=hotpink247129191, thick]
        (axis cs:192,247.776248144333)
        --(axis cs:192,255.652293455667);

        \path [draw=hotpink247129191, thick]
        (axis cs:256,251.147152073492)
        --(axis cs:256,270.025812326508);

        \path [draw=hotpink247129191, thick]
        (axis cs:384,252.752141895489)
        --(axis cs:384,260.153802104511);

        \path [draw=hotpink247129191, thick]
        (axis cs:512,284.893751024468)
        --(axis cs:512,312.544517375532);

        \path [draw=hotpink247129191, thick]
        (axis cs:768,253.591079835156)
        --(axis cs:768,385.316418164844);

        \path [draw=hotpink247129191, thick]
        (axis cs:1024,234.553065881147)
        --(axis cs:1024,608.142295718853);

        \addplot [thick, darkorange2551270, mark=o, mark size=1.5, mark options={solid}]
        table {%
                8 408.9649338
                16 490.7207646
                32 588.2597708
                64 729.8311684
                128 403.71781575
                192 763.4712065
                256 1385.8198738
                384 1642.146252
                512 1681.7073818
                768 1837.17103375
                1024 4471.2618065
            };

        \addplot [thick, steelblue55126184, mark=o, mark size=1.5, mark options={solid}]
        table {%
                8 350.965411
                16 356.1323506
                32 466.6635376
                64 610.5962682
                128 810.3085536
                192 894.1837972
                256 996.5984034
                384 1142.5767528
                512 1263.0058544
                768 1569.340435
                1024 1577.6902244
            };

        \addplot [thick, mediumseagreen7717574, mark=o, mark size=1.5, mark options={solid}]
        table {%
                8 252.9387324
                16 253.0262516
                32 256.7591236
                64 324.0695552
                128 363.9698516
                192 389.8196148
                256 426.65261
                384 487.3836102
                512 563.135935
                768 562.8612714
                1024 801.4380266
            };

        \addplot [thick, hotpink247129191, mark=o, mark size=1.5, mark options={solid}]
        table {%
                8 253.8295834
                16 257.8832338
                32 253.5560404
                64 258.8229006
                128 257.8739642
                192 251.7142708
                256 260.5864822
                384 256.452972
                512 298.7191342
                768 319.453749
                1024 421.3476808
            };
        \addlegendentry{UAM: Sync (Post Voting)}
        \addlegendentry{OAM: VotesBatching}
        \addlegendentry{OAM: VotePropose (RenameGroup)}
        \addlegendentry{UAM: Vote (RenameGroup)}
    \end{axis}

\end{tikzpicture}
        \footnotesize
        \vspace*{-1mm}
        \hspace*{16mm}(c)
        \vspace*{2mm}

        \phantomcaption
        \label{fig:vote-latency}
    \end{subfigure}

    \begin{subfigure}[b]{.3\linewidth}
        \centering
        \begin{tikzpicture}[scale=0.7]

    \definecolor{darkgray176}{RGB}{176,176,176}
    \definecolor{darkorange2551270}{RGB}{255,127,0}
    \definecolor{hotpink247129191}{RGB}{247,129,191}
    \definecolor{lightgray204}{RGB}{204,204,204}
    \definecolor{mediumseagreen7717574}{RGB}{77,175,74}
    \definecolor{steelblue55126184}{RGB}{55,126,184}

    \begin{axis}[
            legend cell align={left},
            legend style={
                    fill opacity=0.8,
                    draw opacity=1,
                    text opacity=1,
                    at={(0.03,0.97)},
                    anchor=north west,
                    draw=lightgray204
                },
            tick align=outside,
            tick pos=left,
            title={Vote Traffic over Different Group Sizes},
            x grid style={darkgray176},
            xlabel={Group Size},
            xmin=-42.8, xmax=1074.8,
            xtick style={color=black},
            y grid style={darkgray176},
            ylabel={Total Traffic (KB)},
            ymin=-294.702229546654, ymax=6268.4462540063,
            ytick style={color=black},
            ymajorgrids=true, grid style=dashed,
        ]
        \path [draw=darkorange2551270, semithick]
        (axis cs:8,54.9058858863553)
        --(axis cs:8,55.5245828636447);

        \path [draw=darkorange2551270, semithick]
        (axis cs:16,101.049499670173)
        --(axis cs:16,102.359484704827);

        \path [draw=darkorange2551270, semithick]
        (axis cs:32,191.68734862294)
        --(axis cs:32,194.96929200206);

        \path [draw=darkorange2551270, semithick]
        (axis cs:64,375.423291537543)
        --(axis cs:64,381.750145962457);

        \path [draw=darkorange2551270, semithick]
        (axis cs:128,744.140575083953)
        --(axis cs:128,750.993409291047);

        \path [draw=darkorange2551270, semithick]
        (axis cs:192,1114.12488311364)
        --(axis cs:192,1124.50714813636);

        \path [draw=darkorange2551270, semithick]
        (axis cs:256,1467.07669929125)
        --(axis cs:256,1493.01119133375);

        \path [draw=darkorange2551270, semithick]
        (axis cs:384,2223.54142898106)
        --(axis cs:384,2225.62068039394);

        \path [draw=darkorange2551270, semithick]
        (axis cs:512,2934.60612215017)
        --(axis cs:512,2987.13098722483);

        \path [draw=darkorange2551270, semithick]
        (axis cs:768,4434.46755761257)
        --(axis cs:768,4463.14416113743);

        \path [draw=darkorange2551270, semithick]
        (axis cs:1024,5857.74195831429)
        --(axis cs:1024,5970.12132293571);

        \path [draw=steelblue55126184, thick]
        (axis cs:8,37.0843230825029)
        --(axis cs:8,38.3121612924971);

        \path [draw=steelblue55126184, thick]
        (axis cs:16,70.384258575641)
        --(axis cs:16,72.871600799359);

        \path [draw=steelblue55126184, thick]
        (axis cs:32,133.349880287614)
        --(axis cs:32,139.952854087386);

        \path [draw=steelblue55126184, thick]
        (axis cs:64,264.352657912536)
        --(axis cs:64,276.732888962464);

        \path [draw=steelblue55126184, thick]
        (axis cs:128,528.966903799703)
        --(axis cs:128,543.192080575297);

        \path [draw=steelblue55126184, thick]
        (axis cs:192,793.407723226585)
        --(axis cs:192,814.969620523415);

        \path [draw=steelblue55126184, thick]
        (axis cs:256,1026.41663243278)
        --(axis cs:256,1077.29977381722);

        \path [draw=steelblue55126184, thick]
        (axis cs:384,1591.50281359132)
        --(axis cs:384,1593.59171765868);

        \path [draw=steelblue55126184, thick]
        (axis cs:512,2064.07335368412)
        --(axis cs:512,2170.21570881588);

        \path [draw=steelblue55126184, thick]
        (axis cs:768,3165.88327537001)
        --(axis cs:768,3224.20695900499);

        \path [draw=steelblue55126184, thick]
        (axis cs:1024,4114.16367699717)
        --(axis cs:1024,4341.23554175283);

        \path [draw=mediumseagreen7717574, thick]
        (axis cs:8,10.2460809895554)
        --(axis cs:8,11.7019658854446);

        \path [draw=mediumseagreen7717574, thick]
        (axis cs:16,13.823979255106)
        --(axis cs:16,16.181098869894);

        \path [draw=mediumseagreen7717574, thick]
        (axis cs:32,16.8740715706581)
        --(axis cs:32,23.3751471793419);

        \path [draw=mediumseagreen7717574, thick]
        (axis cs:64,27.853914234203)
        --(axis cs:64,40.046867015797);

        \path [draw=mediumseagreen7717574, thick]
        (axis cs:128,53.273831148993)
        --(axis cs:128,67.957418851007);

        \path [draw=mediumseagreen7717574, thick]
        (axis cs:192,49.3526219742998)
        --(axis cs:192,98.7262842757002);

        \path [draw=mediumseagreen7717574, thick]
        (axis cs:256,70.7553417025558)
        --(axis cs:256,122.313017672444);

        \path [draw=mediumseagreen7717574, thick]
        (axis cs:384,157.758186465417)
        --(axis cs:384,158.055876034583);

        \path [draw=mediumseagreen7717574, thick]
        (axis cs:512,148.815448303458)
        --(axis cs:512,256.225957946542);

        \path [draw=mediumseagreen7717574, thick]
        (axis cs:768,116.094124431806)
        --(axis cs:768,291.987906818194);

        \path [draw=mediumseagreen7717574, thick]
        (axis cs:1024,280.307004138003)
        --(axis cs:1024,508.264480236997);

        \path [draw=hotpink247129191, semithick]
        (axis cs:8,3.62270152393487)
        --(axis cs:8,3.63276722606513);

        \path [draw=hotpink247129191, semithick]
        (axis cs:16,3.65382191485851)
        --(axis cs:16,3.69461558514149);

        \path [draw=hotpink247129191, semithick]
        (axis cs:32,3.72704945133631)
        --(axis cs:32,3.77724742366369);

        \path [draw=hotpink247129191, semithick]
        (axis cs:64,3.8876677932847)
        --(axis cs:64,3.9181915817153);

        \path [draw=hotpink247129191, semithick]
        (axis cs:128,4.24558511080357)
        --(axis cs:128,4.27511801419643);

        \path [draw=hotpink247129191, semithick]
        (axis cs:192,4.58915592312787)
        --(axis cs:192,4.62295345187213);

        \path [draw=hotpink247129191, semithick]
        (axis cs:256,4.98610910625846)
        --(axis cs:256,5.04084401874154);

        \path [draw=hotpink247129191, semithick]
        (axis cs:384,5.73254534616329)
        --(axis cs:384,5.77722027883671);

        \path [draw=hotpink247129191, semithick]
        (axis cs:512,6.47199595910284)
        --(axis cs:512,6.54441029089716);

        \path [draw=hotpink247129191, semithick]
        (axis cs:768,7.98289583076239)
        --(axis cs:768,8.05147916923761);

        \path [draw=hotpink247129191, semithick]
        (axis cs:1024,9.51691017204139)
        --(axis cs:1024,9.55652732795861);

        \addplot [semithick, darkorange2551270, mark=o, mark size=1.5, mark options={solid}]
        table {%
                8 55.215234375
                16 101.7044921875
                32 193.3283203125
                64 378.58671875
                128 747.5669921875
                192 1119.316015625
                256 1480.0439453125
                384 2224.5810546875
                512 2960.8685546875
                768 4448.805859375
                1024 5913.931640625
            };

        \addplot [thick, steelblue55126184, mark=o, mark size=1.5, mark options={solid}]
        table {%
                8 37.6982421875
                16 71.6279296875
                32 136.6513671875
                64 270.5427734375
                128 536.0794921875
                192 804.188671875
                256 1051.858203125
                384 1592.547265625
                512 2117.14453125
                768 3195.0451171875
                1024 4227.699609375
            };

        \addplot [thick, mediumseagreen7717574, mark=o, mark size=1.5, mark options={solid}]
        table {%
                8 10.9740234375
                16 15.0025390625
                32 20.124609375
                64 33.950390625
                128 60.615625
                192 74.039453125
                256 96.5341796875
                384 157.90703125
                512 202.520703125
                768 204.041015625
                1024 394.2857421875
            };

        \addplot [semithick, hotpink247129191, mark=o, mark size=1.5, mark options={solid}]
        table {%
                8 3.627734375
                16 3.67421875
                32 3.7521484375
                64 3.9029296875
                128 4.2603515625
                192 4.6060546875
                256 5.0134765625
                384 5.7548828125
                512 6.508203125
                768 8.0171875
                1024 9.53671875
            };
        \addlegendentry{UAM: Sync (Post Voting)}
        \addlegendentry{OAM: VotesBatching}
        \addlegendentry{OAM: VotePropose (RenameGroup)}
        \addlegendentry{UAM: Vote (RenameGroup)}
    \end{axis}

\end{tikzpicture}
        \footnotesize
        \hspace*{16mm}(d)
        \vspace*{-2mm}
        \phantomcaption
        \label{fig:vote-bandwidth}
    \end{subfigure}\hfill
    \begin{subfigure}[b]{.3\linewidth}
        \centering
        \input{tikz_DS_16x64.tex}
        \begin{minipage}{.1cm}
            \vfill
        \end{minipage}
        \footnotesize
        \hspace*{16mm}(e)
        \vspace*{-2mm}

        \phantomcaption
        \label{fig:server-16}
    \end{subfigure}\hfill
    \begin{subfigure}[b]{.3\linewidth}
        \centering

        \begin{tikzpicture}[scale=0.7]
    \definecolor{lightgray204}{RGB}{204,204,204}
    \definecolor{mediumseagreen7717574}{RGB}{77,175,74}
    \definecolor{steelblue55126184}{RGB}{55,126,184}
    \definecolor{darkgray176}{RGB}{176,176,176}
    \definecolor{darkorange2551270}{RGB}{255,127,0}
    \definecolor{hotpink247129191}{RGB}{247,129,191}
    \definecolor{lightgreen128247129}{RGB}{128,247,129}

    \definecolor{purple12855126}{RGB}{128,55,126}
    \definecolor{slateblue12877175}{RGB}{128,77,175}
    \begin{axis}[
            legend cell align={left},
            legend style={
                    fill opacity=0.8,
                    draw opacity=1,
                    text opacity=1,
                    at={(0.01,0.99)},
                    anchor=north west,
                    draw=lightgray204,
                    font=\small
                },
            tick align=outside,
            tick pos=left,
            title={Server latency: 16 groups of 64 members},
            xlabel={Number of requests sent by all clients},
            xmin=-409.6, xmax=8601.6,
            xtick style={color=black},
            x grid style={darkgray176},
            y grid style={darkgray176},
            ylabel={Average Latency (Time to complete) (sec)},
            ymin=-10, ymax=300,
            ytick style={color=black},
            ymajorgrids=true, grid style=dashed,
        ]

        \path [draw=purple12855126, draw opacity=0.72156862745098, semithick]
        (axis cs:0,2.85621245517631)
        --(axis cs:0,2.90303394482369);

        \path [draw=purple12855126, draw opacity=0.72156862745098, semithick]
        (axis cs:1024,11.9851864356394)
        --(axis cs:1024,12.1035935643606);

        \path [draw=purple12855126, draw opacity=0.72156862745098, semithick]
        (axis cs:2048,21.0419849859869)
        --(axis cs:2048,21.6145210140131);

        \path [draw=purple12855126, draw opacity=0.72156862745098, semithick]
        (axis cs:4096,48.5400144514387)
        --(axis cs:4096,77.6095643485613);

        \path [draw=purple12855126, draw opacity=0.72156862745098, semithick]
        (axis cs:8192,206.99526443073)
        --(axis cs:8192,268.82533196927);

        \path [draw=steelblue55126184, semithick]
        (axis cs:0,5.73879585162196)
        --(axis cs:0,5.86537294837804);

        \path [draw=steelblue55126184, semithick]
        (axis cs:1024,18.3770904369577)
        --(axis cs:1024,18.4418547630423);

        \path [draw=steelblue55126184, semithick]
        (axis cs:2048,31.5265432616953)
        --(axis cs:2048,32.1169967383046);

        \path [draw=steelblue55126184, semithick]
        (axis cs:4096,67.9372115794349)
        --(axis cs:4096,92.9493568205652);

        \path [draw=steelblue55126184, semithick]
        (axis cs:8192,217.307727218316)
        --(axis cs:8192,280.912052381684);

        \path [draw=none, semithick]
        (axis cs:0,2.83341450589157)
        --(axis cs:0,2.91366309410843);

        \path [draw=none, semithick]
        (axis cs:1024,5.21378794990722)
        --(axis cs:1024,5.32779125009278);

        \path [draw=none, semithick]
        (axis cs:2048,8.96642387836397)
        --(axis cs:2048,9.03608132163603);

        \path [draw=none, semithick]
        (axis cs:4096,20.1130286945689)
        --(axis cs:4096,20.3493417054311);

        \path [draw=none, semithick]
        (axis cs:8192,78.9836452971907)
        --(axis cs:8192,88.6564939028093);

        \path [draw=darkorange2551270, semithick]
        (axis cs:0,5.73402718572176)
        --(axis cs:0,5.81244721427824);

        \path [draw=darkorange2551270, semithick]
        (axis cs:1024,10.5631389460879)
        --(axis cs:1024,10.6569202539121);

        \path [draw=darkorange2551270, semithick]
        (axis cs:2048,16.9784624096547)
        --(axis cs:2048,17.1531611903453);

        \path [draw=darkorange2551270, semithick]
        (axis cs:4096,33.4509306219081)
        --(axis cs:4096,33.8682669780919);

        \path [draw=darkorange2551270, semithick]
        (axis cs:8192,93.0488168962909)
        --(axis cs:8192,111.031491903709);

        \path [draw=slateblue12877175, draw opacity=0.290196078431373, semithick]
        (axis cs:0,2.83955234289687)
        --(axis cs:0,2.93062165710313);

        \path [draw=slateblue12877175, draw opacity=0.290196078431373, semithick]
        (axis cs:1024,3.34572852141678)
        --(axis cs:1024,3.41449187858322);

        \path [draw=slateblue12877175, draw opacity=0.290196078431373, semithick]
        (axis cs:2048,4.35526030765161)
        --(axis cs:2048,4.47054449234839);

        \path [draw=slateblue12877175, draw opacity=0.290196078431373, semithick]
        (axis cs:4096,7.66861615062551)
        --(axis cs:4096,7.81173584937449);

        \path [draw=slateblue12877175, draw opacity=0.290196078431373, semithick]
        (axis cs:8192,16.6711389983742)
        --(axis cs:8192,16.8346646016258);

        \path [draw=mediumseagreen7717574, semithick]
        (axis cs:0,5.72622326873116)
        --(axis cs:0,5.86121233126884);

        \path [draw=mediumseagreen7717574, semithick]
        (axis cs:1024,8.03528599282544)
        --(axis cs:1024,8.08322480717456);

        \path [draw=mediumseagreen7717574, semithick]
        (axis cs:2048,8.72243405184936)
        --(axis cs:2048,8.82025314815064);

        \path [draw=mediumseagreen7717574, semithick]
        (axis cs:4096,14.5678701817652)
        --(axis cs:4096,14.7157302182348);

        \path [draw=mediumseagreen7717574, semithick]
        (axis cs:8192,29.4318980609584)
        --(axis cs:8192,30.3207451390416);

        \path [draw=lightgreen128247129, draw opacity=0.749019607843137, semithick]
        (axis cs:0,2.8475462372364)
        --(axis cs:0,2.8762589627636);

        \path [draw=lightgreen128247129, draw opacity=0.749019607843137, semithick]
        (axis cs:1024,3.18958013566897)
        --(axis cs:1024,3.25518866433103);

        \path [draw=lightgreen128247129, draw opacity=0.749019607843137, semithick]
        (axis cs:2048,3.89980816562983)
        --(axis cs:2048,4.02805023437017);

        \path [draw=lightgreen128247129, draw opacity=0.749019607843137, semithick]
        (axis cs:4096,6.21514563136128)
        --(axis cs:4096,6.27684396863872);

        \path [draw=lightgreen128247129, draw opacity=0.749019607843137, semithick]
        (axis cs:8192,10.7865371096905)
        --(axis cs:8192,11.5356192903095);

        \path [draw=hotpink247129191, semithick]
        (axis cs:0,5.72230112383039)
        --(axis cs:0,5.79625247616961);

        \path [draw=hotpink247129191, semithick]
        (axis cs:1024,8.53564475558531)
        --(axis cs:1024,8.60479444441469);

        \path [draw=hotpink247129191, semithick]
        (axis cs:2048,11.9912431562581)
        --(axis cs:2048,12.131239243742);

        \path [draw=hotpink247129191, semithick]
        (axis cs:4096,20.4727367154777)
        --(axis cs:4096,20.6290936845223);

        \path [draw=hotpink247129191, semithick]
        (axis cs:8192,38.8321219900372)
        --(axis cs:8192,39.7101528099628);

        \addplot [semithick, steelblue55126184, mark=|, mark size=3, mark options={solid}]
        table {%
                0 5.8020844
                1024 18.4094726
                2048 31.82177
                4096 80.4432842
                8192 249.1098898
            };
        \addplot [semithick, steelblue55126184, dotted, mark=|, mark size=3, mark options={solid}]
        table {%
                0 2.8796232
                1024 12.04439
                2048 21.328253
                4096 63.0747894
                8192 237.9102982
            };

        \addplot [semithick, darkorange2551270, mark=|, mark size=3, mark options={solid}]
        table {%
                0 5.7732372
                1024 10.6100296
                2048 17.0658118
                4096 33.6595988
                8192 102.0401544
            };
        \addplot [semithick, darkorange2551270, dotted, mark=|, mark size=3, mark options={solid}]
        table {%
                0 2.8735388
                1024 5.2707896
                2048 9.0012526
                4096 20.2311852
                8192 83.8200696
            };

        \addplot [semithick, mediumseagreen7717574, mark=|, mark size=3, mark options={solid}]
        table {%
                0 5.7937178
                1024 8.0592554
                2048 8.7713436
                4096 14.6418002
                8192 29.8763216
            };
        \addplot [semithick, mediumseagreen7717574, dotted, mark=|, mark size=3, mark options={solid}]
        table {%
                0 2.885087
                1024 3.3801102
                2048 4.4129024
                4096 7.740176
                8192 16.7529018
            };

        \addplot [semithick, hotpink247129191, mark=|, mark size=3, mark options={solid}]
        table {%
                0 5.7592768
                1024 8.5702196
                2048 12.0612412
                4096 20.5509152
                8192 39.2711374
            };
        \addplot [semithick, hotpink247129191, dotted, mark=|, mark size=3, mark options={solid}]
        table {%
                0 2.8619026
                1024 3.2223844
                2048 3.9639292
                4096 6.2459948
                8192 11.1610782
            };

        \legend{Send+Sync All OAM, Send Only All OAM,  Send+Sync 50\% OAM,Send Only 50\% OAM, Send+Sync 90\% UAM, Send Only 90\% UAM, Send+Sync All UAM, Send Only All UAM}
    \end{axis}
\end{tikzpicture}
        \footnotesize
        \hspace*{16mm}(f)
        \vspace*{-2mm}
        \phantomcaption
        \label{fig:server-512}
    \end{subfigure}
    \caption{Experimental evaluation of \sysname. Results, averaged over 5
        trials, include standard deviation as error bars. In (a)(b), data is
        from the last starting client for multi-client operations. Only (e)(f)
        feature data from both servers and clients in US-East, instead of
        cross-regional -- this was an optimization to allow for faster server benchmarks.}

    \label{fig:experiments}
\end{figure*}

We note that, as far as we are aware, ours is the first comprehensive
end-to-end benchmark of a messaging system built on MLS. Hence, these
absolute performance numbers may be of independent interest.

\paragraph{Experimental setup}
We perform our experimental evaluations in a networked setting. We use AWS
EC2 to run our AS and DS (on a m7g.medium in US-West-2), and our client machines (in US-East-2, 8 clients per t4g.small instance), in order to test in the WAN setting, with the only exception of server processing time analysis (where all instances are in US-East-2). See Appendix \ref{addl_exp} for additional details.

\paragraph{Microbenchmarks} We present both client latency and client-to-server bandwidth metrics for various operations in \figsref{fig:micro-latency}{fig:micro-bandwidth} across group sizes ranging from 8 to 1024. Latency is measured from the initiation of a client request to the end of processing of server responses, possibly containing new messages from other members. This latency breakdown includes request and its corresponding sync request generation, network communication, and processing.

Most message types, when compared to \baseline, bear extra bandwidth overheads due to digital signatures, each adding an average of 882.20 bytes on average across 5 trials of all group sizes. The impact on latency and bandwidth varies with group size, as illustrated in \figref{fig:micro-latency}. %

Governance state in group additions depends on group size, growing approximately linearly. This state is only relayed in a GroupStateAnnouncement message when adding users and is depicted in \figref{fig:micro-bandwidth}.

A significant portion of end-to-end latency is dominated by network channel establishment and data
transmission, accounting for $99.6\%\sim 99.7\%$ for UAM and $68.8\%\sim
    99.0\%$ for OAM, appeared as Network Overhead in \figref{fig:netTimes}.
\sysname latency is mostly negligible, with exceptions
being the addition of many members or syncing large message quantities in
large groups. OAM bears a significant overhead than UAMs and incurs a
longer generation and processing time. This is because OAMs are sent
through commit messages, which by default carry out key rotations for
forward secrecy and post compromise recovery. Our evaluations involve a
worst-case configuration of an MLS group in which these updates are linear
in size. Governance state announcements increase in size linearly in group
size because the RBAC, which is part of the governance state, tracks
information for each group member.

For a group of 64, a text message's transmission requires $258$ ms and
$3.54$ KiB of bandwidth. Meanwhile, adding all 63 members takes $485$ ms
and incurs a bandwidth of $177$ KiB. These metrics are tabulated in
\figref{fig:netTimes}. Notably, costs for sending text messages remain
constant, but costs scale linearly for group rename actions and governance
state announcements, largely due to key rotations, request metadata, and
RBAC-related data.

\paragraph{Voting macro-benchmark}
To assess the performance of a complex governance procedure, we benchmarked a representative voting workload. For groups ranging from 16 to 1024 members, we measured latency and bandwidth (averaged over 5 trials, with standard deviations) for a vote to rename the group, reporting the results in \figsref{fig:vote-latency}{fig:vote-bandwidth}. In our tests, one user starts the vote, and all members cast votes simultaneously via unordered messages. Once a client receives enough votes, it batches them in an ordered message.

We noted a linear rise in time and bandwidth (communication complexity of sent messages). For a group of 1024, it takes about $0.50\,(0.17)$ seconds (standard deviation in parentheses) for all to vote and $1.58\,(0.35)$ seconds for a member to batch and commit votes with an OAM. Our data shows operations use $9.53\,(0.02)$ KB network traffic per voter and $4227.70\,(113.53)$ KB for the batching member, making our voting approach viable for large groups.

\paragraph{Server evaluation}
We evaluate how well our system implementation scales with different request loads by measuring how fast our delivery service can handle requests. The latency is defined as the timespan from the start of the first client's request to the end of the last client request.

We test 1024 users with various workloads: 100\% unordered messages with
21-character strings, 100\% ordered rename messages of length $11\pm 2$
characters (we have the group names depend on the client name, which can vary in size), and randomly generated mixed workloads with either half or 90\%
unordered requests, with the remainder as rename requests.

We vary the total number of requests for these four
different workloads and measure total latency provided by our
server and the achieved throughput. The offered workload ranges from 0 up to $2^{17}$ requests, capped at around 250 seconds. We performed the same benchmark in a setup with 512 and 16 groups of sizes 2 (direct messaging) and 64, both involving 1024 clients (see Appendix \ref{addl_exp:serverload}).

We report on our results in \figsref{fig:server-16}{fig:server-512}. In situations with ordering contention, when others' valid OAM arrives at the DS before a client's generated OAM does, the client needs multiple messages and communication rounds for its request completion. Larger group size means more read per request and also higher contention likelihood, which resulted in a slower completion.

Yet, even under challenging conditions where all OAMs are within large
groups, our system can handle an average of 32.89 incoming requests
(accompanied by over 2,072 message retrievals) every second. In a more
typical scenario with 90\% UAM and 10\% OAM, the server processes a minimum
of 127 requests (8192 message retrievals) per second for groups of 64, and
585.14 requests per second for groups of 2. Employing a more powerful
server could decrease the message retrieval latency, but its efficacy may
diminish with a high volume of ordered messages per group.

\section{Conclusion}
This paper introduces the novel goal of private hierarchical governance for
encrypted group messaging. We show how community moderation systems widely
used on plaintext platforms can be adapted to the E2EE setting while
maintaining privacy, integrity, and accountability. Our solution is a
radical departure from prior E2EE moderation approaches which focus on
platform-driven moderation. As a result, private hierarchical governance
opens up new possibilities for abuse mitigation that do not suffer from the
transparency and accountability issues that arise with platform-driven
solutions.

Our design pushes the execution of governance to client devices and makes
use of the messaging layer to maintain shared encrypted state. Instead of
focusing on hard-coding specific policies, our design enables a framework
for expressing general policies, such as voting and content filter
enforcement. Through enabling reporting to both platform and community
moderators, our design provides channels to inform moderation at both
levels. We conduct a security analysis of our design by reasoning through
possible attack scenarios. We build and benchmark a prototype encrypted
messaging platform that realizes private hierarchical
governance in order to demonstrate its practicality.

\section*{Acknowledgements}

\noindent This work was funded in part by NSF grant CNS-2120651. We thank
the reviewers and our shepherd for their helpful feedback.

\bibliographystyle{plain}
\bibliography{refs}

\appendices
\section{MLS Background}
\label{sec:mls-background}
In this section, we summarize details about the MLS protocol that
are relevant to our work.

\paragraph{MLS architecture}
MLS~\cite{mls-architecture} relies on two services: an
\textit{authentication service (AS)} for managing user identities and a
\textit{delivery service (DS)}  for transporting messages. The concrete
design of these services are not specified, and implementors are free to
design within the API~\cite{mls-protocol}.

The AS is a service that serves the role of a traditional PKI, mapping user
identities (usernames) to certificates. Looking ahead, we use credentials
that include for each user a long-lived digital signature public key. This
long-lived public key can be used to verify the authenticity of further
cryptographic keys, and ultimately allows cryptographically verifying, for
example, sent messages as emanating from a particular sender username. As
in any PKI, security relies on the AS being a trusted third-party that
provides users with the most up-to-date views of these mappings. We suggest
that deployments use a key transparency
mechanism~\cite{CONIKS,SEEMless,VeRSA,Mog} to enable users or auditors to
check for malicious behavior on the part of the authentication service.

The DS transfers (encrypted) messages between users in the network. In
contrast to solutions built out of independent pairwise channels, in our
implementation we opt for a server fan-out design. Clients send messages to
the delivery service along with a list of intended recipients. The DS then
forwards the messages along to those recipients. Over the course of its
operation, the delivery service does not need to keep track of who belongs
to which group, however, it can infer membership based on which recipients
a message specifies.

\paragraph{Protocol overview}  The MLS protocol~\cite{mls-protocol}
provides a mechanism for group encrypted messaging. To do so, it uses the
TreeKEM protocol~\cite{treekem,AsyncRatchetingTree} to allow a group to
efficiently maintain a shared secret that evolves over time. KEM (key
encapsulation mechanism) public keys are authenticated via long-lived
signing keys that are, in turn, authenticated by the AS. TreeKEM allows for
efficient changes to group membership and provides strong forward secrecy
and post-compromise security guarantees.

Protocol messages in MLS can be either public or private. Public MLS
messages are signed by the sender. Private MLS messages are signed by the
sender and then encrypted using a current group-held symmetric key with an
appropriate authenticated encryption with associated data (AEAD) scheme
(such as AES-GCM~\cite{aesgcm}). Group members can verify the sender of
both private and public messages; public messages can additionally be
verified by the DS should that be useful to applications.

MLS has two classes of messages, \emph{application messages} and
\emph{handshake messages}. The former are private messages used to transmit
plaintext data to the group. Delivery of application messages is best-effort, and MLS is explicitly designed to allow message reordering or even
dropping of messages. Handshake messages are more complicated as they are
used to maintain shared group state, such as the current TreeKEM. To add a
new client to a group, the user adding the new client prepares a welcome
message. These include a serialization of the current shared cryptographic
state, and are sent as a private message.

The shared group state must evolve consistently over time, as users join
and leave, with updates to user KEM public keys for forward secrecy, etc.
MLS therefore requires a consensus mechanism to ensure that clients agree
on this evolution.

\section{Additional Security Analysis}
\label{sec:add-security}

Our design prevents the platform from directly learning about governance
actions, messages, the group name/topic, block list, votes cast, election
outcomes, and member profiles. However, the platform may be able to infer
via traffic analysis who serves as the moderator of the group, the outcome
of a vote, and the type of governance messages sent in the group. Whether
such attacks will be effective in practice is unclear, given that all
messages are encrypted and the possibility of deploying countermeasures
that have been explored in other contexts such as
TLS~\cite{trafficAnalysis}.

To elaborate, suppose a group has a policy in which only moderators can add
new members. Although \texttt{Add} messages are encrypted, the DS may still
infer which messages add new members through keeping track of the recipient
list of messages. For instance, the DS may notice that after $U_1$ sent a
message (whose contents are encrypted), the recipient list in $U_1$'s next
message contains one more user, $U_2$. As a result, the DS can infer that
the first encrypted message probably added $U_2$ to the group and that
$U_1$ is authorized to add new members to the group. These inferences are
likely to work in some settings and not others (e.g., due to noise), and so
future work will be needed to evaluate their practicality.  What's more,
our approach to governance is amenable to deployment of traffic analysis
mitigations such as padding, dummy messages, and metadata private
messaging.

Only received action messages can be reported with cryptographic assurance
to the moderator, others will only be trustworthy should client software be
honest. To elaborate, in a conversation between Alice and Bob, Alice can
report the messages she receives from Bob but cannot cryptographically
prove that Bob received particular messages from her. This limitation also
exists for other asymmetric cryptographic message franking
solutions~\cite{AMF}. The reason is that nothing prevents a malicious
reporter with modified client software from generating a new action
message, signing it, and reporting it to a moderator---without actually
sending the action message to any group.

Another related limitation is that ordering information of messages is not
currently cryptographically verifiable, and only trustworthy should client
software be honest. In fact there is no absolute ordering of content
actions, since these are encoded as UAMs, and so this issue is in some
sense fundamental. That said, one might add DS-signed time stamps to
(encrypted) messages to enforce some partial ordering.

Our reporting mechanisms do not immediately enable the reporting of
deviation from protocol behavior. For instance, a malicious client sending
an honest client an incorrect initial state is not reportable. Even though
the incorrect group state announcement is reportable, proving that it is
incorrect would require reporting every commit sent in the group in order
to compute what the correct group state should be. This is in general not
practical or desirable. Future work could explore providing cryptographic
assurance for commit sequences using zero-knowledge proofs. Regardless,
honest clients can still issue claims to the platform moderator that a
client generated an incorrect group state announcement. If many such
reports are received, say from a majority of a group, a platform moderator
would have reason to take action against the reported inviter.

MLS provides strong forward-secrecy (FS) and post
compromise security (PCS). FS entails the confidentiality of messages
sent before a compromise occurs and PCS guarantees the confidentiality
of messages sent after a healing procedure following a compromise. We
now discuss how our addition of governance, in particular, our
mechanism for shared encrypted state, interacts with the FS and PCS
properties of MLS. Recall that state updates are sent via encrypted
commit messages and that new users learn the current aggregate state
via a message sent by a current group member. These update messages are
protected by FS and PCS, but the reliance of our system on maintaining
long-term aggregate state changes the implications of FS. In
particular, when a group state update message (sent upon a new user
joining a group) is compromised, the attacker can infer the contents of
prior messages that led to the aggregate state they observed. This
issue surrounding FS seems inherent to similar systems that must
maintain long-term accumulated state, such as end-to-end encrypted
backups. Semantic information from compromised messages can also leak
information about prior messages sent, even if those messages are
cryptographically protected by FS. Finally, given that users are often
not required to turn on disappearing messages, device compromise
trivially reveals to an attacker past messages in the clear, which are
stored on-device.

In summary, our governance mechanism requires maintaining long-term
encrypted state that may reveal information about messages sent before
FS ratchets. The extent of such leakage is highly dependent on what
specific policies are deployed. Furthermore, we cannot deploy a
``disappearing messages'' type feature for governance state in a
straightforward manner. If our aim were to hide information about past
governance messages, we could do so by re-setting governance state
(like the list of moderators) periodically, but this would likely be
prohibitive from a usability standpoint. To be clear, the addition of
governance does not impact forward secrecy of content-carrying
messages. Rather, maintaining long-term state that gradually evolves
over the lifetime of a group necessitates that prior governance
messages be reflected within the current governance state.

\section{Additional Implementation Details}
\label{addl_impl}
\paragraph{Communications and Persistent States} We designed and
implemented a custom protocol for communication between clients and the two servers. To do so we use standard approaches, and used
JSON for data serialization and
web sockets for transferring messages between clients and servers.

\paragraph{Signature} We use the
ed25519-dalek \cite{dalek-link}
implementation of ed25519 for governance digital signatures. Clients sign all actions and include the signature in the request.

\paragraph{Unordered and Ordered Voting} We implemented both voting in all
ordered and unordered (except for the start and ending
batch-commit message(s) being ordered) format. When all clients start to
vote at the same time, ordered voting can have very high contention that
even if coupled with an exponential back-off mechanism (a common retry
strategy), the completion time could still last for minutes for bigger
groups. Unordered voting on the other hand brings voting time down to
seconds, but unordered messages are relayed best-effort and could be lost,
with a remedy that the client can batch-commit their unordered vote
themselves should their vote does not appear in other members' batch-commit.

\section{Additional Experimental Results and Details}
\label{addl_exp}

\begin{figure*}[p]
    \centering
    \setlength\tabcolsep{4pt} %
    \scriptsize %
    \begin{tabular}{lccccc|cccc}
        \toprule
        \multicolumn{6}{c|}{\textbf{Operation  Latency \& Traffic}} & \multicolumn{4}{c}{\textbf{Sync  Latency \& Traffic}}                                                                                                                                                                                                                                                                                                   \\
        \textbf{Action}                                             & \textbf{Request}                                      & \textbf{Network}                 & \textbf{Post-}                 & \textbf{Total }                   & \textbf{Traffic}                 & \textbf{Network}                   & \textbf{Message}                      & \textbf{Total }                      & \textbf{Traffic}                 \\
                                                                    & \textbf{Gen.}                                         & \textbf{Delay}                   & \textbf{Processing}            & \textbf{Latency}                  &                                  & \textbf{Delay}                     & \textbf{Processing}                   & \textbf{Latency}                     &                                  \\
                                                                    & (ms)                                                  & (ms)                             & (ms)                           & (ms)                              & (KB)                             & (ms)                               & (ms)                                  & (ms)                                 & (KB)                             \\
        \midrule
        Invite (1023 invitees)                                      & 66.94\textsuperscript{(1.09)}                         & 1170.27\textsuperscript{(90.75)} & 200.25\textsuperscript{(0.17)} & 1866.37\textsuperscript{(103.33)} & 10254.38\textsuperscript{(2.85)} &                                    &                                       &                                      &                                  \\
        Add (1023 invitees)                                         & 337.32\textsuperscript{(0.75)}                        & 551.04\textsuperscript{(25.3)}   & 1.3\textsuperscript{(0.09)}    & 1283.11\textsuperscript{(39.56)}  & 2752.14\textsuperscript{(0.9)}   & 3563.57\textsuperscript{(1948.93)} & 20.10\textsuperscript{(0.83)}         & 3585.59\textsuperscript{(1542.01)}   & 1795.32\textsuperscript{(0.47)}  \\
        GovStateAnnoun.                                             & 1.95\textsuperscript{(0.07)}                          & 298.87\textsuperscript{(22.03)}  & 0.53\textsuperscript{(0.04)}   & 363.62\textsuperscript{(1.94)}    & 72.12\textsuperscript{(0.11)}    &                                    &                                       &                                      &                                  \\
        \midrule
        Accept                                                      & 3.84\textsuperscript{(0.24)}                          & 254.45\textsuperscript{(3.67)}   & 0.07\textsuperscript{(0.07)}   & 259.12\textsuperscript{(3.36)}    & 9.26\textsuperscript{(0.02)}     & 2535.59\textsuperscript{(1149.73)} & 614428.41\textsuperscript{(51687.51)} & 616969.1\textsuperscript{(51930.98)} & 3529.61\textsuperscript{(0.64)}  \\
        \midrule
        RenameGroup                                                 & 141.37\textsuperscript{(1.09)}                        & 644.03\textsuperscript{(82.48)}  & 0.5\textsuperscript{(0.02)}    & 937.78\textsuperscript{(82.64)}   & 600.5\textsuperscript{(0.16)}    & 1476.81\textsuperscript{(266.87)}  & 8.21\textsuperscript{(2.59)}          & 1486.52\textsuperscript{(248.27)}    & 748.49\textsuperscript{(0.15)}   \\
        \midrule
        VotePropose (Rename)                                        & 94.4\textsuperscript{(29.66)}                         & 574.26\textsuperscript{(74.46)}  & 0.85\textsuperscript{(0.05)}   & 801.44\textsuperscript{(114.64)}  & 394.29\textsuperscript{(113.98)} & 1242.37\textsuperscript{(358.08)}  & 6.27\textsuperscript{(2.45)}          & 1250.14\textsuperscript{(264.72)}    & 645.38\textsuperscript{(56.95)}  \\
        \midrule
        Vote (Rename)                                               & 0.63\textsuperscript{(0.01)}                          & 547.78\textsuperscript{(183.57)} & 0.05\textsuperscript{(0.0)}    & 549.16\textsuperscript{(172.16)}  & 9.53\textsuperscript{(0.02)}     & 3606.64\textsuperscript{(620.86)}  & 602.40\textsuperscript{(26.95)}       & 4214.05\textsuperscript{(447.57)}    & 5913.89\textsuperscript{(56.18)} \\

        \midrule
        Send (10-char Text)                                         & 0.72\textsuperscript{(0.03)}                          & 271.94\textsuperscript{(1.28)}   & 0.08\textsuperscript{(0.01)}   & 273.83\textsuperscript{(1.24)}    & 9.17\textsuperscript{(0.02)}     & 613.35\textsuperscript{(43.62)}    & 0.62\textsuperscript{(0.04)}          & 615.59\textsuperscript{(40.18)}      & 454.1\textsuperscript{(0.08)}    \\
        \midrule
        Send (100-char Text)                                        & 0.65\textsuperscript{(0.03)}                          & 257.31\textsuperscript{(0.62)}   & 0.08\textsuperscript{(0.0)}    & 259.01\textsuperscript{(0.54)}    & 9.47\textsuperscript{(0.02)}     & 593.48\textsuperscript{(9.72)}     & 0.65\textsuperscript{(0.02)}          & 595.85\textsuperscript{(9.37)}       & 454.38\textsuperscript{(0.09)}   \\

        \bottomrule
    \end{tabular}
    \caption{Operation latency breakdown for a user in a group of 1024 users in \sysname (5-trial average (std.)). Clients are on US-East AWS instances while DS/AS are on US-West. For operations requested by multiple clients simultaneously, data from the 513th starting client is used. Total latency represents the time between loading the pre-operation group state and saving the post-operation group state, including server processing and key package generation/update delays.}
    \label{fig:netTimes1024}
\end{figure*}

\paragraph{Additional results} We include additional experimental results
for our system in \figref{fig:netTimes1024}, which reports micro-benchmark
results for a group consisting of 1024 members.

\paragraph{Client Instances} To mimic performance on a low-budget device,
every 8 clients run on a t4g.small instance, which has 2 GiB of RAM and 2~vCPU,
and network bandwidth up to 5 Gbps.

\paragraph{Server Instance}
The AS and DS run on a single m7g.medium instance, which has 4 GiB of RAM and 1~vCPU, and network bandwidth up to 12.5 Gbps.

\paragraph{Text Field Length} Text message
content consists of strings of $10\pm 1$ (default unless specified otherwise) or $100$ characters. Rename message content contains strings of $15\pm 1$ characters. Vote message content contains the string ``yes''.
Invites add a single new user with a name consisting of $1\sim 4$ characters to the group.

\paragraph{Benchmarking Tooling} We use boto3 with SSH (Paramiko) to automate the process of creating and running instances for experiments. We use an t4g.xlarge instance to issue the operation command to all instances, which can reach 1024 clients (128 instances) within 3 seconds unless the commands are too demanding and are draining instance resources.

\paragraph{Server Evaluation Detail}
\label{addl_exp:serverload}
The workload is generated by each client looping through requests consisted
of the target distribution, with all clients starts uniformly in the loop. We
disable group state saving to reduce client-induced overheads. To mimic
natural interactions, users send sync messages before their requests in these
workloads.

\section{Pseudocode Specification}
\label{ap:pseudocode}
We present a pseudocode specification of our governance protocol in
\figref{fig:pseudocode}. In particular, we demonstrate how clients create
and process messages within the system, including user reports. We indicate
explicitly where our system calls out to MLS. The
$\mlsApi.\sendUnorderedMsg$ call corresponds to sending an MLS
\texttt{ApplicationMessage}. The $\mlsApi.\sendOrderedMsg$ corresponds to
usage of our new ordered application message proposal type. An overview of
the MLS-level functionalities can be found in the MLS protocol
specification \cite{mls-protocol} and OpenMLS documentation \cite{openmls}.
We use the variable $\mlsSt$ to denote state associated with the MLS
messaging layer. The variable $\govSt$ represents shared governance state
within a group. The local state a client maintains is represented by
$\localSt$. In addition to storing the shared governance state, $\localSt$
contains the content state $\conSt$, which may differ between users of the
same group, due to a lack of ordering guarantees for content-carrying
messages. Groups have associated identifiers, denoted as $\gid$. The
signature key pair $(\reportPk, \reportSk)$ enables reporting
signatures on messages.

\begin{figure*}[p]
        \centering
        \fhpagess{0.48}{0.48}{ %
                \procedurev{$\initClient(u) \to \localSt, \mlsSt$}\\
                $\mlsSt \getsr \mlsApi.\initClient()$\\
                $\reportPk, \reportSk \getsr \digSig.\keygen()$; store $\reportSk$
                in $\localSt$\\
                Post $\reportPk$ and KeyPackages for user $u$ to AS\\
                $\creturn (\localSt, \mlsSt)$\\

                \procedurev{$\createGroup(\localSt, \mlsSt) \to (\mlsSt',
                                \localSt', \gid)$}\\
                Initialize default $\govSt$\\
                Initialize empty content state $\conSt$\\
                Add entries $(\gid, \govSt)$ and $(\gid, \conSt)$ to $\localSt$\\
                $\creturn (\mlsSt, \localSt)$\\

                \procedurev{$\sendContentMsg(\mlsSt, \localSt, m, \gid) \to (c, \mlsSt',
                                \localSt')$}\\
                $\signature \gets \digSig.\signMsg(\reportSk, m)$\\
                Append $m$ to $\conSt$ for $\gid$\\
                $\creturn \mlsApi.\sendUnorderedMsg(\mlsSt, (m, \signature), \gid),
                        \localSt$\\

                \procedurev{$\sendGovMsg(\mlsSt, \localSt, m, \gid) \to (c, \mlsSt',
                                \localSt')$}\\
                $\signature \gets \digSig.\signMsg(\reportSk, m)$\\
                Append $m$ to $\conSt$ for $\gid$\\
                $\creturn \mlsApi.\sendOrderedMsg(\mlsSt, (m, \signature), \gid), \localSt$\\

                \procedurev{$\sendReport(\mlsSt, \localSt, m, \gid) \to c $}\\
                Retrieve reporting signature $\signature$ for $m$\\
                Append $m$ to $\conSt$ for $\gid$\\
                $\creturn \mlsApi.\sendUnorderedMsg(\mlsSt, (\texttt{Report}, m, \signature),
                        \gid)$\\
        }{
                \procedurev{$\recvMsg(\mlsSt, \localSt, c, \gid) \to (\mlsSt',
                                \localSt')$}\\
                $m, \mlsSt \gets \mlsApi.\recv(\mlsSt, c, \gid)$\\
                Retrieve entries $(\gid, \govSt)$, $(\gid, \conSt)$ from
                $\localSt$\\
                $\govSt, \conSt \gets \exec(\policy, m, \govSt, \conSt)$\\
                Update entries $(\gid, \govSt)$, $(\gid, \conSt)$\\
                $\creturn (\mlsSt, \localSt)$\\

                \procedurev{$\accept(\mlsSt, \localSt, \welcome, \cGov) \to
                                (\mlsSt', \localSt', c)$}\\
                $\mlsSt, \gid \gets \mlsApi.\joinGroup(\mlsSt, \welcome)$\\
                $\govSt \gets \mlsApi.\recv(\mlsSt, \cGov, \gid)$\\
                Initialize empty content state $\conSt$\\
                Store entry $(\gid, \govSt)$ and $(\gid, \conSt)$ to $\localSt$\\
                $m \gets (\texttt{Accept,}, H(\govSt))$\\
                $c \getsr \mlsApi.\sendUnorderedMsg(\mlsSt, (m,
                        \digSig.\signMsg(\reportSk, m)), \gid)$\\
                $\creturn (\mlsSt, \localSt, c)$\\

                \procedurev{$\verifReport(\mlsSt, \localSt, c, \gid) \to b$}\\
                $(\texttt{Report}, m, \signature) \gets \mlsApi.\recv(\mlsSt, c,
                        \gid)$\\
                $\creturn \digSig.\verify(\reportPk, m, \signature)$\\

                \procedurev{$\mlsApi.\sendOrderedMsg(\mlsSt, m, \gid) \to
                                \mlsSt'$}\\
                $\creturn \mlsApi.\sendCommit(\mlsSt, \ordAppMsgProp(m, \gid))$

        }
        \caption{A pseudocode specification of our governance protocol.}
        \label{fig:pseudocode}
\end{figure*}

\newpage
\section{Meta-Review}

The following meta-review was prepared by the program committee for the 2024
IEEE Symposium on Security and Privacy (S\&P) as part of the review process as
detailed in the call for papers.

\subsection{Summary}
This paper brings the ability for a group or community on an end-to-end encrypted platform to have moderators and rules that are internally enforced while maintaining the ability for the platform itself to have ultimate authority over moderation decisions. The proposed solution builds on top of the consensus mechanisms included in MLS. The authors use the existing MLS channels plus independent governance public keys associated to each group member to propose actions, and RBAC to determine if the group member is allowed to take such an action. Additionally, group members maintain governance state to ensure an attacker cannot fork governance actions by group members. The authors implement their proposal and report on the performance of their implementation. Their experiments indicate that their proposal scales reasonably well.

\subsection{Scientific Contributions}
\begin{itemize}
    \item Addresses a Long-Known Issue
    \item Provides a Valuable Step Forward in an Established Field
    \item Establishes a New Research Direction
\end{itemize}

\subsection{Reasons for Acceptance}
\begin{enumerate}
    \item The paper shows how to overcome a key impediment to large-scale deployments of encrypted social media. Governance for end-to-end encrypted messaging is a long known issue and this paper advances the discussion.

    \item The implementation and evaluation demonstrate feasibility of the solution.
          This paper creates an open source implementation built on top of MLS that
          will be useful in understanding the capabilities that MLS can provide for
          governance actions
\end{enumerate}

\subsection{Noteworthy Concerns}
While the paper considers various attack scenarios in its security evaluation, it does not formally analyze its contributions.

\end{document}